\documentclass[runningheads
]{llncs}
\usepackage{graphicx}
\graphicspath{{gfx/},{experiments/},}  

\RequirePackage[nolist]{acronym}

\usepackage{subcaption}

\usepackage{listings,algorithm}
\usepackage[noend]{algpseudocode}
\algdef{SE}[DOWHILE]{Do}{doWhile}{\algorithmicdo}[1]{\algorithmicwhile\ #1}%

\usepackage{amsmath, amssymb}

\usepackage[table,dvipsnames]{xcolor}

\usepackage{enumitem}

\newcolumntype{x}[1]{>{\centering\arraybackslash\hspace{0pt}}p{#1}}

\usepackage[nospace]{cite}

\usepackage{xfrac,xspace}
\usepackage{relsize}
\usepackage[normalem]{ulem}

\usepackage{wrapfig}

\usepackage{tikz}
\usetikzlibrary{patterns}
\usetikzlibrary{positioning}
\usetikzlibrary{arrows,arrows.meta,bending}
\usetikzlibrary{fit}
\tikzset{
  treenode/.style = {align=center, inner sep=0pt, text centered},
  basis/.style = {
    pattern=north east lines,
    pattern color=lightgray,
  },
  newbasis/.style={
    pattern=north west lines,
    pattern color=teal!80!black!60!white,
  },
  other/.style={
    pattern=north east lines,
    pattern color=red,
  },
  frontier/.style={
    pattern=north east lines,
    pattern color=yellow!80!black,
  },
  pw/.style={
    pattern=north east lines,
    pattern color=green!80!black,
  },
  q0class/.style={
    pattern=north east lines,
    pattern color=red!40!white,
  },
  q1class/.style={
    pattern=crosshatch dots,
    pattern color=blue!60!white,
  },
  q2class/.style={
    pattern=north west lines,
    pattern color=green!50!black!60!white,
  },
  basic/.style = {
    fill=white,
  }
}
\usepackage{pgfplots}
\pgfplotsset{compat=1.8}
\pgfplotsset{vasymptote/.style={
    before end axis/.append code={
        \draw[densely dashed] ({rel axis cs:0,0} -| {axis cs:#1,0})
        -- ({rel axis cs:0,1} -| {axis cs:#1,0});
    }
}}

\usepackage[outline]{contour}
\contourlength{1.2pt}

\usepackage{mdframed}

\usepackage{comment}

\usepackage{longtable}
\usepackage{pifont}
\newcommand{\xmark}{\ding{55}}

\RequirePackage[%
  bookmarks,
  unicode,
  breaklinks=true,
  ]{hyperref}
  
\usepackage[nameinlink]{cleveref}

\usepackage{todonotes}
\usepackage{booktabs}

\newcommand{\splitatcommas}[1]{%
	\begingroup
	\begingroup\lccode`~=`, \lowercase{\endgroup
		\edef~{\mathchar\the\mathcode`, \penalty0 \noexpand\hspace{0pt plus 1em}}%
	}\mathcode`,="8000 #1%
	\endgroup
}

\makeatletter
\renewcommand{\paragraph}{\@startsection{paragraph}{5}{0em}%
  {.7ex plus .2ex minus .1ex}%
  {-.5em}%
  {\bfseries}}
\makeatother

\makeatletter
\def\orcidID#1{\smash{\href{http://orcid.org/#1}{\protect\raisebox{-1.25pt}{\protect\includegraphics{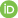}}}}}
\makeatother

\allowdisplaybreaks[4]

\Crefname{equation}{eq.}{eqs.}
\crefname{equation}{equation}{equations}
\Crefname{figure}{Fig.}{Figs.}
\crefname{figure}{figure}{figures}
\Crefname{tabular}{Table}{Tables}
\crefname{tabular}{table}{tables}
\Crefname{definition}{Def.}{Defs.}
\crefname{definition}{definition}{definitions}
\Crefname{proposition}{Prop.}{Props.}
\crefname{proposition}{proposition}{propositions}
\Crefname{section}{Sec.}{Sections}
\crefname{section}{section}{sections}
\Crefname{subsection}{Sec.}{Sections}
\crefname{subsection}{subsection}{subsections}
\crefname{algorithm}{algorithm}{algorithms}
\crefname{listing}{code}{code\ blocks}
\crefformat{footnote}{#2\footnotemark[#1]#3}

\newcommand{\myparagraph}[1]{\smallskip\noindent \emph{#1}}

\newcommand{\io}{\mathop{\sim_\textnormal{IO}}}
\newcommand{\bio}{\mathop{\approx_\textnormal{IO}}}
\newcommand{\notio}{\mathop{\not\sim_\textnormal{IO}}}
\newcommand{\ido}{\mathop{\sim_\textnormal{IDO}}}
\newcommand{\notido}{\mathop{\not\sim_\textnormal{IDO}}}

\DeclareMathAlphabet{\mathsc}{OT1}{cmr}{m}{sc}

\mathchardef\mhyph="2D 

\newcommand{\tto}[1]{\stackrel{#1}{\to}}

\newcommand{\hide}[1]{}

\colorlet{colcodekey}{PineGreen!60!green!70!black}
\colorlet{colcodeidentifier}{Sepia}
\colorlet{colcodenumeric}{blue!25!black}
\colorlet{colcodesyncaction}{violet!90}
\colorlet{colcodecomment}{Black!60}
\colorlet{coldistribution}{Cerulean!60!Black}
\lstdefinelanguage{Kepler}{  
  keywords={},  
  keywords=[2]{toplevel,and,or,pand,vot,fdep,spare,be,sbe,rbox,seq,csp,wsp,hsp},
  keywords=[3]{fail,repair,dorm,prio,fcfs,rand,lambda,prob,res,rate,mean},
  keywords=[4]{exponential,erlang,uniform,normal,lognormal,%
               weibull,gamma,rayleigh,fsteq,dirac,%
               exp,uni,dir},
  otherkeywords={},
  morecomment=**[l]{//},      
  morecomment=**[s]{/*}{*/},
  moredelim=**[is][\color{colcodenumeric}]{^}{^},     
  moredelim=**[is][\color{colcodekey}]{-|}{|-},       
  moredelim=**[is][\slshape]{__}{__},                 
}
\lstdefinestyle{Kepler}{
  language=Kepler,
  xleftmargin=2em,
  basewidth=0.5em,
  basicstyle={\scriptsize\ttfamily},
  identifierstyle=\color{colcodeidentifier},
  keywordstyle=\bfseries,
  keywordstyle=[2]\color{colcodesyncaction},
  keywordstyle=[3]\bfseries\color{colcodekey},
  keywordstyle=[4]\bfseries\color{coldistribution},
  commentstyle=\color{colcodecomment},
  stringstyle=\mdseries,
  showstringspaces=false,
  numbers=left,
  numberstyle=\scriptsize\ttfamily\color{colcodecomment},
  numbersep=0.9em,
  tabsize=4,
  frame=none,
  aboveskip=\bigskipamount,
  belowskip=\medskipamount,
  abovecaptionskip=\smallskipamount,
  belowcaptionskip=\smallskipamount,
  captionpos=b,
  escapeinside={`}{`},
  mathescape=true,  
}

\begin{acronym}[ABCDEFGHIJK]
    \acro{MM}{Mealy machine}
    \acro{MDM}{Mealy delay machine}
    \acro{AAL}{active automata learning}
    \acro{SUL}{system under learning}
    \acro{MAT}{Minimal Adequate Teacher}
    \acro{OQ}{output query}
    \acro{EQ}{equivalence query}
    \acro{EO}{equivalence oracle}
    \acro{CDF}{cumulative distribution function}
    \acro{PDF}{probability density function}
    \acro{I/O}{input/output}
    \acro{RPP}{Rural Postman Problem}
    \acro{CPP}{Chinese Postman Problem}
    \acro{MDP}{Markov decision processes}
    \acro{MM}{Mealy machine}
    \acro{MDM}{Mealy delay machine}
    \acro{w.l.o.g.}{without loss of generality}
    \acro{MLE}{maximum likelihood estimation}
    \acro{DKW}{Dvoretzky–Kiefer–Wolfowitz}
    \acro{CTMC}{continous-time Markov chain}
    \acro{RV}{random variable}
    \acroplural{OQ}[OQs]{output queries}
    \acroplural{EQ}[EQs]{equivalence queries}
    \acroplural{MDP}[MDPs]{Markov decision processes}
    \acroplural{MM}[MMs]{Mealy machines}
    \acroplural{MDM}[MDMs]{Mealy delay machines}
    \acroplural{I/O}[I/O]{inputs/outputs}
\end{acronym}

\begin{document}

\title{Automata Learning -- Expect Delays!\thanks{%
This work was partially funded by
DFG grant 389792660 as part of \href{https://perspicuous-computing.science}{TRR~248 CPEC},
the European Union (EU) under the INTERREG North Sea project STORM\_SAFE of the European Regional Development Fund,
and the EU's Horizon 2020 research and innovation programme under Marie Skłodowska-Curie grant agreement 101008233 (MISSION).
}}
 \author{Gabriel~Dengler \orcidID{0000-0002-4217-4952} \and Sven Apel \orcidID{0000-0003-3687-2233} \and Holger Hermanns \orcidID{0000-0002-2766-9615}}
 \authorrunning{G. Dengler et al.}

\institute{Saarland University, Saarland Informatics Campus, Saarbrücken, Germany
}

\maketitle

\begin{abstract}
    This paper studies \ac{AAL} in the presence of stochastic delays. We consider Mealy machines that have stochastic delays associated with each transition and explore how the learner can efficiently arrive at faithful estimates of those machines, the precision of which crucially relies on repetitive sampling of transition delays. While it is possible to na\"ively integrate the delay sampling into \ac{AAL} algorithms such as $L^*$, this leads to considerable oversampling near the root of the state space. We address this problem by separating conceptually the learning of behavior and delays such that the learner uses the information gained while learning the logical behavior to arrive at efficient input sequences for collecting the needed delay samples. We put emphasis on treating cases in which identical input/output behaviors might stem from distinct delay characteristics.
    Finally, we provide empirical evidence that our method outperforms the na\"ive baseline across a wide range of benchmarks and investigate its applicability in a realistic setting by studying the join order in a relational database.
\end{abstract}

\acresetall

\section{Introduction}
\label{sec:introduction}

The goal of \ac{AAL} is to infer the behavioral structure of a \ac{SUL} by executing actions upon the \ac{SUL} and observing its resulting behavior~\cite{learning-practical-perspective, AAL-overview}. 
This technique is the basis for inferring formal models of possibly large and complex systems for the sake of further analysis~\cite{black-box-checking, black-box-checking-learnlib, falsification-aal, prob-black-box-checking}. 
Besides the logical behavior of a system, an important aspect in real-word systems is their real-time behavior, which often comes with stochastic imprecisions due to communication delays or internal computation steps consuming time. 
Learning the timing behavior can become very relevant, for instance to detect bottlenecks in the implementation or to adjust to observed system performance.
One of the prominent past applications of \ac{AAL} is related to finding critical bugs in protocol implementations~\cite{prognosis-network-aal}. Since these implementations first and foremost
are meant to provide high performance communication services, it seems natural to consider information about the delay characteristics of the various steps as additional valuable information to discover fast and slow steps of the system and how these could be accounted for. These delay characteristics usually need to be considered as being stochastic in nature and are representable as manifestations of continuous-time probability distributions.

There is a plethora of basic model families that come to mind as target models~\cite{automata-zoo} for \ac{AAL} in such a setting, including timed automata~\cite{alur-timed-automata,larsen-timed-automata}, probabilistic automata~\cite{DBLP:journals/njc/SegalaL95,probabilistic-automata} including Markov decision processes, Markov automata~\cite{yuxin,prob-automata-cont-time}, and stochastic (timed) automata~\cite{dargenio-stochastic-automata, stochastic-timed-automata-bbb}. Indeed, inspiring research work on \ac{AAL} has been undertaken in many facets for timed automata~\cite{timed-automata-myhill-nerode, oneclock-timed-automata, real-time-automata, nondet-real-time-automata, active-learning-timed-unobs-resets, time-delay-mealy, learning-delay-rta, learn-mealy-with-timers, learn-local-timers, learn-multiclock-timed-reset-info, learning-mealy-machines-timers-symbolic} as well as \acp{MDP}~\cite{learning-stochastic-reactive, l*-based-mdp-learning, robust-anytime-mdps, PAC-statistical-checking-MDPs, aal-mdps-baum-welch}. However, to the best of our knowledge, automata models that provide explicit support for stochastic delays have not been put into the \ac{AAL} research focus. At the same time, these appear as natural candidates in scenarios where the delays of activities follow continuous probability distributions over time. These can be represented directly within stochastic (timed) automata or approximated as precisely as needed within Markov automata via phase-type fitting~\cite{asmussen:1996:fitting}. 

This paper aims at exploring this topic from ground up. For this purpose, we start with the goal of performing black-box learning of a deterministic~\ac{MM} while collecting the time needed to traverse transitions alongside learning. While we assume determinism for the state space, the timing behavior is assumed to follow continuous probability distributions. This makes it necessary to traverse each transition multiple times to obtain statistically sound estimations. We will argue that it is theoretically justified and algorithmically beneficial to work with a global constant $k$ of necessary transition traversals per transition to be considered. Furthermore, we consider it natural (albeit challenging) that the teacher has full knowledge of the time-abstract behavior, but that the time dependencies are entirely outside their knowledge, and thus the ground truth needs to (and can only) be inferred approximately through observations.

\myparagraph{Related work.} For models that exhibit probabilistic decisions associated to actions, e.g., \acp{MDP} or probabilistic \acp{MM}, it is---as in our scenario---essential that each input query will be asked multiple times to get statistically robust estimates of the \ac{SUL}'s behavior. Concepts from the literature that adapt $L^*$ for probabilistic models either introduce queries that return the sampled results as an observation tree~\cite{learning-stochastic-reactive} or equip the teacher with the ability to sample the \ac{SUL} and thereby obtain estimations of the transition probabilities to be used by the learner~\cite{l*-based-mdp-learning}.
Thereby, the learner needs to reset the system frequently, which leads to considerable oversampling near the root, whereas other parts of the system are covered to a much lesser extent.
Instead of actively learning the \ac{MDP}, there is also the possibility to adapt passive learning approaches like \textsc{Alergia}~\cite{carrasco:1994:alergia} to a probabilistic setting~\cite{mao:2012:ioalergia-mdps}. However, experiments demonstrate an overall better performance for active learning approaches \cite{l*-based-mdp-learning}.

In a timed setting, delays induced by transitions are often modeled using clocks or timers in automata or \acp{MM}:
This can be explicitly achieved by splitting transitions into two consecutive transitions~\cite{DBLP:conf/cav/NicollinS91,learn-mealy-with-timers}, where time is spent in the intermediate state, i.e., by waiting for the expiration of a delay timer. While these models are (distributions aside) more general than the \ac{MM} structure considered in this paper, their use in \ac{AAL} research comes with other side constraints that hamper application in the real-world settings we focus on: They predominantly consider deterministic behavior for each clock/timer and sometimes assume the existence of a smart teacher \cite{timed-automata-myhill-nerode, learn-multiclock-timed-reset-info} that can explicitly answer more complex queries or provide hints to the learner. We do not assume the existence of any of these helpful features. In the literature, there are also some suggestions to learn stochastic timed models passively \cite{pedro:2012:learn-sta-sample-executions, mediouni:2017:improved-sta-learning, sen:2004:learning-sta-merging}, then resorting to state merging as in \textsc{Alergia}~\cite{carrasco:1994:alergia}. These approaches do not incorporate efficient input sequences to ensure a uniform coverage of the state space and provide no guarantees regarding the inferred logical structure.

\myparagraph{Contributions.} To the best of our knowledge, this paper is the first to establish a formal framework for the active learning of automata with stochastic delays. It starts off by extending the concept of standard \acp{MM} (\Cref{sec:background}) and introduces \acp{MDM} 
(\Cref{sec:delay-mealy-machine}) that describe the probabilistic behavior of transitions delays with continuous probability distributions.
We first consider models where the learned logical structure (an \ac{MM}) is the one underlying the \ac{MDM} (\Cref{sec:methodology-min-mealy}). 
We then turn to the more intricate situation where certain sequences through the learned \ac{MM} come with distinct delay characteristics albeit being indistinguishable from the perspective of the logical structure in isolation (\Cref{sec:methodology-state-sep}). Our theoretical findings are supported by a prototypical implementation, which we empirically evaluate on publicly available benchmark models (\Cref{sec:evaluation}) and on a database application inspired by real-world contexts~(\Cref{sec:practice}).

\section{Preliminaries}
\label{sec:background}

We recall relevant notions regarding \acp{MM} and \ac{AAL}, and also discuss some basics of continuous probability distributions and their estimation. 

\paragraph{\aclp{MM}.}
\label{def:mealy-machines}
An \ac{MM} $\mathcal{M}$ is a 6-tuple $(S, I, O, s_0, \delta, \lambda)$, in which $S$ is a finite set of states, $I$ the inputs, $O$ the outputs, $s_0 \in S$ the initial state, $\delta : S\times I \rightarrow S$ the transition function, and $\lambda : S\times I \rightarrow O$ the output function.
 
 We denote with $|\mathcal{M}|$ the number of states of $\mathcal{M}$. 
 A state $s\in S$ is a \emph{sink} iff for all $i\in I$, $\delta(s, i) = s$. 
 As usual, transition and output functions are extended to input words of length $n$ by the functions $\delta: S \times I^n \rightarrow S$ and $\lambda: S\times I^n \rightarrow O^n$.
 We use $w_j$ to identify the $j$-th character of the word $w \in I^*$. 
 We use $\delta(w)$ and $\lambda(w)$ to refer to $\delta(s_0, w)$, respectively $\lambda(s_0, w)$. 
  As common in the \ac{AAL} context, we silently assume that all states in $S$ can be reached from $s_0$ by some word in $I^*$. Furthermore, we can characterize the equivalence behavior of an \ac{MM} in terms of the \acp{I/O} of its states as follows. 
\begin{definition}
    If given a set of input sequences $W \subset I^*$ from an \ac{MM} $\mathcal{M} = (S, I, O, s_0, \delta, \lambda)$, we say that two states $s$ and $t$ from $S$ are $W$-equivalent, denoted $s \equiv^{_{W}} t$, iff $\lambda(s, w) = \lambda(t, w)$ for all $w\in W$. 
    
   They are \ac{I/O}-equivalent, denoted $s \io t$, iff $s \equiv^{_{I^*}} t$.
\end{definition}
   
The induced relation $\io\subseteq S\times S$ is indeed an equivalence relation on the states of~$\mathcal{M}$. It can be lifted to an equivalence relation on \acp{MM}: Two \acp{MM} are equivalent if their initial states are equivalent in an MM spanned by their disjoint union (whatever state is fixed as initial state in that MM).

Since we are dealing with deterministic structures, an alternative  way of characterizing \ac{I/O}-equivalence of two \acp{MM} harvests the notion of \emph{bisimulation}~\cite{park71}.
\begin{definition}
    \label{def:bisimulation-mealy}
    Two \acp{MM} over the same input set $I$, $\mathcal{M} = (\splitatcommas{S, I, O, s_{0}, \delta, \lambda})$ and $\mathcal{M}' = (\splitatcommas{S', I, O, s'_{0}, \delta', \lambda'})$,  are said to be bisimilar, denoted $\mathcal{M} \bio \mathcal{M}'$, iff there is a relation $R \subseteq S \times S'$ (named bisimulation relation) such that $s_{0}~R~s'_{0}$, and for each pair $(s,s') \in  S\times S'$, $s~R~s'$ implies for all inputs $i\in I$ that 
 $\lambda(s, i) = \lambda'(s', i)$ (\ac{I/O} equality), and $\delta(s, i)~R~\delta'(s', i)$ (state transition correspondence).
 \end{definition}
\begin{lemma}\label{lem:io}
Two \acp{MM} are \ac{I/O}-equivalent if and only if they are bisimilar.
\end{lemma}
An \ac{MM} $\mathcal{M}$ is called minimal, iff no \ac{MM}  bisimilar to $\mathcal{M}$ is of smaller size, i.e., $|\mathcal{M}| \leq |\mathcal{M}'|$ holds for any bisimilar $\mathcal{M}'$.

The minimal \ac{MM} is unique up to permutation of states.
\hide{A bisimulation relation $R$ between bisimilar Mealy machines $\mathcal{M}$ and $\mathcal{M}'$ is minimal, if there is no other bisimulation relation $R' \subset R$ that relates $\mathcal{M}$ and $\mathcal{M}'$.}

\paragraph{Active automata learning.} The goal of \ac{AAL} is to infer the structure of an \ac{SUL}, in our case an \ac{MM} $\mathcal{S} = (S, I, O, s_0, \delta_{\mathcal{S}}, \lambda_{\mathcal{S}})$, by using \acp{OQ} and \acp{EQ}. An \ac{OQ} for an input word $w$ returns the output word $\lambda_S(w)$. An \ac{EQ} checks whether a hypothesis \ac{MM} $\mathcal{H} = (H, I, O, h_0, \delta_{\mathcal{H}}, \lambda_{\mathcal{H}})$ is bisimilar to $\mathcal{S}$, and, if not, returns a \emph{counterexample} $w$ so that $ \lambda_{\mathcal{S}}(w) \neq \lambda_{\mathcal{H}}(w)$.

\myparagraph{Data structure.} In $L^*$-based algorithms~\cite{Lstar}, the learner maintains as data structure a set of access words $Q \subseteq I^*$ to reach each state in $H$ of the current hypothesis $\mathcal{H}$ and a set of test words $T \subseteq I^*$ to distinguish each state, also called the test set. More precisely, $Q$ and $T$ have to satisfy the following properties:
\begin{itemize}
    \item \emph{Separability:} No two distinct states in $\{ \delta_{\mathcal{S}}(w) \mid w\in Q\}$ are $T$-equivalent.
    \item \emph{Closedness:} For every $q\in Q$ and $i\in I$, there is some $q' \in Q$ such that $\delta_{\mathcal{S}}(qi) \equiv^{_{T}} \delta_{\mathcal{S}}(q')$.
\end{itemize}
To make statements about these properties efficiently, the learner saves the outputs of $\lambda_{\mathcal{S}}(w)$ for the words $w \in (Q \cup \{qi \mid q\in Q, i\in I\}) \times T$.
By construction, the sets $Q$ and $T$ stay separable during $L^*$, but the closedness property has to be retained by extending the set $Q$ with the state accessible by $qi$ if no suitable counterpart $q'$ exists in $Q$ already. Once both separability and closedness are achieved, an \ac{EQ} is issued to the teacher, leading either to the confirmation of the hypothesis $\mathcal{H}$ or an extension of the sets $Q$ and $T$.

\myparagraph{Equivalence oracles.} In practice, the teacher is often unable to answer \acp{EQ} by a so-called \emph{perfect} \ac{EO}.
Instead, the teacher usually resorts to some \emph{test suite} that executes a set of test words and either finds a counterexample and otherwise assumes the system's equivalence.

Under reasonable assumptions on the \ac{SUL}, it is possible to construct a test suite to prove that, after successfully executing all test inputs, the learned system must be equivalent to the \ac{SUL}.
A prominent method to construct such a test suite practically is the $W$-method~\cite{W-method, W-method-2} which assumes a maximum number $e$ of extra states. However, its complexity grows exponentially with the selection of $e$. Therefore, practitioners often make additional real-world motivated assumptions on the \ac{SUL} (e.g., some inputs always leading to a sink, multiple global phases like login or logout, sets of inputs used together, working with a known bug-free model, etc.) to reduce the size of the test suite \cite{small-test-suites, vaandrager:2024:ka-completeness}.
Alternatively, one can also resort to randomized \acp{EO}, but they do only offer probabilistic guarantees (we refer the reader to \cite{mohri2018foundations, AALpy} for an overview).

\paragraph{Continuous probability distributions.}
A continuous \ac{CDF}  $F$ with non-ne\-ga\-tive support describes the probability that the \ac{RV} $X$ with support $\mathbb{R}_0^{+}$ is smaller than $x$, i.e., $F(x)=\Pr (X \leq x)$.
The set of all corresponding \acp{CDF} is denoted as $\mathcal{F}^+_0$.

\myparagraph{Approximating \acp{CDF} from samples.} We will later be interested in reconstructing an approximand $\widehat{F}$ from its samples $X_1, ..., X_k$.  
The most direct approximand is the empirical \ac{CDF} ${\widehat {F}}_{k}(x)={\frac{1}{k}}\sum_{i=1}^{k}\mathbf {1}_{X_{i}\leq x}$ ($\mathbf{1}_{A}$ is the indicator of event $A$), which counts for each value $x$ the percentage of observed samples smaller than or equal to $x$. As this representation is usually unhandy, researchers have developed methods working with simpler representations instead. A common approach is to assume a distribution class, e.g., exponential distributions (for which it is enough to compute the mean of the data points), or the more general class of phase-type distributions. For deriving the best-fitting phase-type distribution given the data points, an abundant number of sophisticated phase-type fitting (or moment-matching) algorithms exist in the
literature~\cite{asmussen:1996:fitting,feldmann:1998:fitting,horvath:2002:phfit,khayari:2003:fitting,thummler:2006:novel,buchholz:2019:online,efficient-phase-type-fitting,johnson:1989:matching,bobbio:2005:matching,horvath:2007:matching,telek:2007:minimal,horvath:2013:moment}.

\myparagraph{Accuracy of approximands.} All approximation methods come
with their accuracy analysis that invariably depends on the number of samples $k$. As the approach we present here is not bound to any specific approximation method, we review the question of approximation accuracy from the foundational perspective.

In full generality, the \ac{DKW} inequality~\cite{dkw-1, dkw-2} provides the probability that a given maximum difference $\varepsilon$ is overshot:
$\Pr {(}\sup _{x\in \mathbb {R}_0^+ }|\widehat{F}_{k}(x)-F(x)|>\varepsilon {)}\leq 2e^{-2k\varepsilon ^{2}}$.
If we are instead interested in bounding the approximation error of the distribution's mean, we can resort to Chebyshev's inequality~\cite{feller:1991:probability-introduction} (adapted to the sample mean) when assuming a maximum variance or Hoeffding's inequality~\cite{hoeffding:1963} when assuming a lower and upper bound to find an appropriate $k$ for an absolute error bound.
Furthermore, in the special case of exponential distributions, we can directly determine the number of samples $k$ needed to achieve a relative error bound (see \Cref{sec:bounding-relative-error-exponential}).

\myparagraph{Checking equality of distributions.} There are many established tests to decide whether two independent sets of samples are likely to come from the same distribution or not~\cite{rubner:1998:earth-mover,two-sample-ks-test,compare-distributions-2}. However, it is not guaranteed that such a test will retain the transitivity property, e.g. when it yields by sampling for three \acp{CDF} $F_1, F_2, F_3$ that $F_1 = F_2$ and $F_2 = F_3$, it may not yield $F_1 = F_3$.
Therefore, researchers have adapted clustering approaches originally developed for multidimensional vectors like $k$-means~\cite{macqueen:1967:k-means} to histograms~\cite{nielsen:2014:alpha-divergence} and empirical \acp{CDF}~\cite{henderson:2015:ep-means}.

\section{Mealy Delay Machines}
\label{sec:delay-mealy-machine}

We now extend the definition of regular \acp{MM} by including  for each transition a description of the stochastic delay behavior.\footnote{In this model, time advances while taking transitions, but it is easily converted into one where time advances in states (at the price of adding intermediate states).} We consider the  
setting where every input triggers an unavoidable delay until the output arrives (in contrast to some timed models~\cite{time-delay-mealy}, where timers can be cancelled by other actions).
\begin{definition}
    \label{def:mealy-delay-machine}
    An \ac{MDM} is a 7-tuple $\mathcal{D} = (\splitatcommas{S, I, O, s_0, \delta, \lambda, D})$, where the first 6 com\-po\-nents constitute an \ac{MM} (called the \emph{underlying} \ac{MM}) and $D : S\times I \rightarrow \mathcal{F}^+_0$ represents the probabilistic delay behavior of each transition as \ac{CDF}.
\end{definition}
As before, $D$ can be extended to inputs of length $n$, resulting in $D : S\times I^n \rightarrow (\mathcal{F}^+_0)^n$. 
Importantly, we make the assumption that the learner will not be able to see the entire \ac{CDF} upon probing the function $D(s, i)$, but, in contrast to the interaction via $\lambda(s, i)$, sees only one sampled value from the \ac{CDF} upon taking the transition.
Formally, the semantic interpretation of an \ac{MDM} is as follows. Any given input word $w$ induces a probability space on timed output words given by a standard cylinder set construction~\cite{DBLP:conf/birthday/BaierHHK08,DBLP:conf/papm/LopezHK01}, detailed in \Cref{sec:sema-mdm}.

We extend the notion of equivalence to the delay case in the obvious way:
\begin{definition}
    \label{def:delay-equivalence}
    If given a set of input sequences $W \subset I^*$ from an \ac{MDM} $\mathcal{D} = (S, I, O, s_0, \delta, \lambda, D)$, we say that two states $s$ and $t$ from $S$ are $W$-equivalent, denoted $s \sim^{_{W}} t$, iff $\lambda(s, w) = \lambda(t, w)$ and $D(s, w) = D(t, w)$ for all $w\in W$.
    
   They are delay-equivalent, denoted $s \ido t$, iff $s \sim^{_{I^*}} t$, 
   \end{definition}
Just as for $\io$, delay equivalence $\ido$ can be lifted to an equivalence relation on \acp{MDM}.
The size of an \ac{MDM} $|\mathcal{D}|$ and its minimality are defined as for \acp{MM}. Note that in the above definition we work with perfect equivalence of \acp{CDF} (because this is the foundational yardstick), but for the practical sampling context, we will work with sampled data points obtained from the \acp{CDF} at hand, and thereby need to resort to a test to check if two distributions are the same. This, in turn, will need to work on approximands obtained from the data points (see~\Cref{sec:background}).

\myparagraph{Problem statement.}
So, we are facing the situation that every transition should be visited sufficiently often to obtain statistically sound estimates of the delay distributions on each transition.
As discussed above, the number $k$ of samples to collect is the decisive parameter governing the accuracy of the approximands obtained. In absence of any other information regarding the concrete \ac{SUL}, we consider it natural to fix this $k$ as a constant across the entire structure to learn.
With this, we can specify the paper's challenge as follows:
\begin{mdframed}[linewidth=0.75pt] 
We are given an \ac{MDM} $\mathcal{D} = (\splitatcommas{S, I, O, s_0, \delta, \lambda, D})$. To interact with $\mathcal{D}$, we can, for any $w\in I^*$, get the outputs from $\lambda(w)$ and one sample from each distribution $D(\delta(w_1...w_{j-1}), w_j)$ for each $j\in \{1, ..., |w|\}$. 
Additionally, we have an \ac{EO} for the underlying \ac{MM} $\mathcal{M}$.
We want to learn the structure of $\mathcal{D}$ and find approximands of the \acp{CDF} of transitions according to $D$ by visiting each transition $\langle s, i\rangle \in S\times I$ at least $k\in \mathbb{N}$ times.
\end{mdframed}
This is the central problem we are aiming to solve. The methods we develop will equally work in settings where individual $k$ are needed for some substructures.

Since two states might be equivalent according to $\io$, but not when considering delays in $\ido$, a peculiar observation when studying delays in an \ac{MM} is:
\begin{corollary}
    The \ac{MM} underlying a minimal \ac{MDM} is not necessarily minimal.
\end{corollary}
Though not necessarily a realistic phenomenon, a minimal \ac{MDM} can in principle be to any order more complex than its corresponding minimal \ac{MM}. In practical scenarios, it is assumable that either (i) the structure of the \ac{MDM} indeed is equivalent to the structure of the underlying minimal \ac{MM} or (ii) some deviation from the minimal \ac{MM} is plausible, but only up to a certain degree. \Cref{sec:methodology-min-mealy} and \Cref{sec:methodology-state-sep} will cover each case separately.

\section{Methodology for Minimal Mealy Machine}
\label{sec:methodology-min-mealy}

As a first challenge, we assume that for an \ac{SUL} \ac{MDM} $\mathcal{D}$ its underlying \ac{MM} $\mathcal{M}$ is indeed of minimal size. Thus, we only have to ensure that every transition is traversed at least $k$ times.

\subsection{Sampling Alongside Active Automata Learning}
\label{sec:sampling-aal}
A straightforward approach is to repeat the sampling of any input word $w\in Q\cup \{ qi \mid q\in Q, i\in I\}$ (see \Cref{sec:background}) $k$ times and to store the resulting delay values, e.g., in an observation tree. This is very close in spirit to what is applied in the learning of probabilistic models such as \acp{MDP}~\cite{learning-stochastic-reactive}. Counterexamples can be handled as usual in $L^*$, although newly discovered test words only need to be traversed once as their delay behavior is not of interest.
Additionally, we can incorporate a cache to ensure that further executions of an input word $w$ are suppressed once the transitions of $w$ have been sampled often enough.

Although this method can correctly infer the structure of the \ac{MDM} $\mathcal{D}$, it has the fundamental problem that a high number of resets is needed, thereby causing many transitions to be traversed an excessive number of times.
This problem also exists for learning regular \acp{MM}. Nonetheless, it is worsened here as each transition needs to be covered multiple times.

\subsection{Efficient Coverage of Learned Mealy Machine}
\label{sec:efficient-coverage-mealy}
To address this problem, we separate the process of inferring the \ac{I/O} behavior from that of sampling the delays.
We start by learning the underlying minimal \ac{MM} $\mathcal{M}'$ underlying $\mathcal{D}$ by some conventional \ac{AAL} algorithms, e.g., $L^*$.
While learning, we can already keep track of delay samples, yet without making any conceptual changes to the algorithmic procedure of the learning algorithm.

After that, we can assume that each transition $\langle s, i\rangle \in S \times I$ of $\mathcal{D}$ has already been traversed some $K(s, i) \in \mathbb{N}_0$ times while inferring the minimal \ac{MM} $\mathcal{M}'$, and we thus know that it needs no more than further $K'(s, i) = \max(0, k - K(s, i))$ visits. With this, we will construct a sequence of input actions to cover every transition in $\mathcal{D}$ sufficiently often, with minimum total cost.

Concretely, the resulting problem can be reduced to the directed \ac{RPP} \cite{rural-postman-problem-1, rural-postman-problem-2} with multiple edges, a variant of the \ac{CPP}~\cite{chinese-postman-problem}. The goal is to find a circuit of minimal cost when some edges need to be traversed a specific number often, while other edges can be used or omitted. As we can reset the \ac{SUL} at any point, we need to incorporate edges from any state to the root state in the graph. Precisely, we resort to a directed multigraph $G = (V, E, m)$, where $V = S$ and the edges $E$ as well as their multiplicity $m$ are defined as:
\begin{align*}
    E &= \{ \langle s, \delta(s, i)\rangle \mid \forall s \in S, \forall i\in I\} \cup \{ \langle s, s_0 \rangle \mid \forall s \in S, s \neq s_0 \} \\
    &m(\langle s, t\rangle) = \sum_{\forall i\in I, \delta(s, i) = t} K'(s, i)\qquad \forall \langle s, t\rangle \in E
\end{align*}
Mandatory edges are those where $m(\langle s, t\rangle) \neq 0$. 
If $G$ is strongly connected and for each node the number of incoming edges (including its multiplicity) is equal to the number of outgoing edges, we can construct an Eulerian
circuit~\cite{graph-theory} in linear time~\cite{chinese-postman-problem} and obtain the optimal strategy to cover all edges as needed.
Since these conditions usually do not apply to $G$, we instead will have to traverse some edges from $E$ more often than strictly required. Unfortunately, this problem is known to be NP-complete in the general case \cite{rural-postman-problem-1, rural-postman-problem-2}.\footnote{We do not know whether the \ac{RPP} limited to the case-specific graphs $G$ we construct is NP-complete, also considering the expansion of the minimal \ac{MM} $\mathcal{M}'$ in \Cref{sec:extended-efficient-coverage-mealy}. The constructed multigraph $G$ is not fully general, because of the multiplicity of the edges and the structure itself being dependent on the previously learned graph and the expanded MM.}
However, when $G$ restricted to mandatory edges forms a weakly connected component (which happens frequently if the minimum number of transition visits $k$ is large), we can compute the additional required edges with minimal cost as follows:
\begin{itemize}
    \item For each node $s$, we calculate the difference of the number between outgoing and incoming edges: $d(s) = \sum_{t \in S, \langle s, t\rangle\in E} m(\langle s, t\rangle) - \sum_{t \in S, \langle t, s\rangle\in E} m(\langle t, s\rangle)$
    \item After that, we solve the min-cost flow problem~\cite{min-cost-flow, min-cost-flow-impl} on the graph induced by the node flow differences as specified in $d(s)$. The solver increases the multiplicity of each edge in $G$ such that $d(s)$ becomes zero for each node $s$ and we obtain minimal costs: For each edge, we can specify how expensive an additional traversal of this edge will be. In our experiments, we assume a constant cost of $1$, but also context-specific costs are readily supported.
\end{itemize}
In the general case, when the subgraph of $G$ spanned by mandatory edges is not weakly connected, it can happen that $G$ including the edges needed for additional traversals does not form a strongly connected component. Then, no Eulerian circuit can be found even if the number of incoming and outgoing edge multiplicities at each node were equal.
In this case, our implementation resorts to the following:
We compute the minimum spanning arborescence of minimum weight when using weight zero for edges that need to be covered and weight one elsewhere~\cite{shortest-arborescence-1, shortest-arborescence-2}, resulting in a weakly-connected graph.

\section{Expansion and State Separation}
\label{sec:methodology-state-sep}

In this section, we study the case that the \ac{MDM} $\mathcal{D}$ has a more complex structure than its underlying minimal \ac{MM} $\mathcal{M}'$. Here, the separation of states is crucial not only to detect different delays for the same logical behavior based on the transition history, but also essential to obtain faithful estimates of distributions.

A general solution to this problem can become costly very rapidly, in case the structure of the \ac{MDM} $\mathcal{D}$ differs significantly from the minimal \ac{MM} $\mathcal{M}'$ or when delay distributions only differ slightly. Therefore, we first focus on practically motivated cases for which we harvest concepts from \Cref{sec:methodology-min-mealy} to handle them efficiently. Later on, we will show that our methods do generalize beyond these cases and can indeed solve the general setting.
We will discuss this as a corollary of the specific findings detailed below~(\Cref{thm:feasible-inj-bisim}).

\begin{wrapfigure}[11]{r}{0.2\textwidth}
    \centering
    \vspace{-25pt}
    \begin{tikzpicture}
    \node[draw, circle, q0class] (q0) at (0, 0) {\contour{white}{$s'_0$}};
    \node[draw, circle, q1class] (q1) at (0, -1.5) {\contour{white}{$s'_1$}};
    \node[draw, circle, q2class] (q2) at (0, -3.0) {\contour{white}{$s'_2$}};
    \node[left=0.3cm of q0] (start) {};

    \path[->,>=stealth',every node/.style={font=\sffamily\scriptsize}]
        (q0) edge[] node[right, align=center] {$a/A$\\ $b/B$} (q1)
        (q1) edge[] node[right, align=center] {$a/B$\\ $b/C$} (q2)
        (q2) edge[bend left=35] node[left, align=center] {$a/C$\\ $b/A$} (q0)
        (start) edge[] (q0)
        ;
\end{tikzpicture}
    \caption{Minimal \ac{MM} $\mathcal{M}'$}
    \label{fig:example1_min}
\end{wrapfigure}
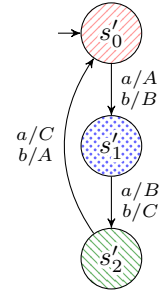

\myparagraph{The useful case.} To make the analysis approachable, we propose restrictions on what the delay behavior induces compared to the \ac{I/O} behavior alone.
The first one states that self-loops in the underlying minimal \ac{MM} indicate that no changes to the system are made. This prevents transitions between distinct states of $\mathcal{D}$ that have the same \ac{I/O} behavior, which is why we use the term \emph{stutter-free}:

\begin{definition}
    \label{def:no-self-loops}
    An \ac{MDM} $\mathcal{D} = (S, I, O, s_0, \delta, \lambda, D)$ is called stut\-ter-free iff for any state $s\in S$ and input $i\in I$ with $\delta(s, i) \io s$, it holds that $\delta(s, i) = s$.
\end{definition}
The second restriction requires that two states which are equivalent regarding the \ac{I/O} behavior also show the same delay behavior at the latest after $d$ steps:
\begin{definition}
    \label{def:d-step-confluence}
    An \ac{MDM} $\mathcal{D} = (\splitatcommas{S, I, O, s_0, \delta, \lambda, D})$ is $d$-step confluent iff for all states $s_1, s_2 \in S$ with $s_1 \io s_2$ and all words $w \in I^{\geq d}$ such that $\delta(s_1, w)$ as well as $\delta(s_2, w)$ do not include a self-loop, it holds $\delta(s_1, w) = \delta(s_2, w)$.
\end{definition}

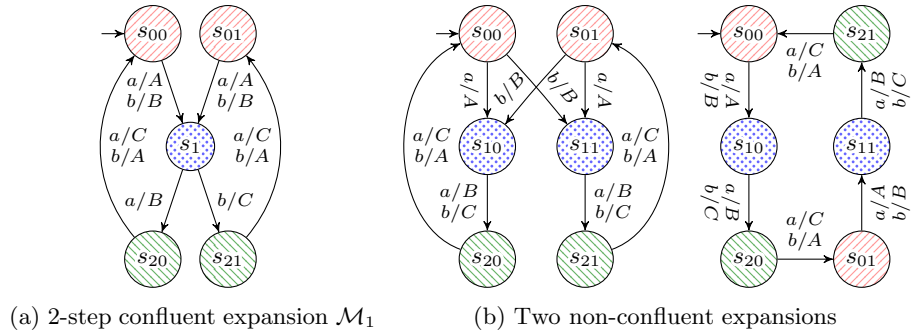
\begin{figure}
    \centering
    \begin{subfigure}{0.4\linewidth}
        \centering
        \begin{tikzpicture}
    \node[draw, circle, q0class] (q00) at (-0.5, 0) {\contour{white}{$s_{00}$}};
    \node[draw, circle, q0class] (q01) at (0.5, 0) {\contour{white}{$s_{01}$}};
    \node[draw, circle, q1class] (q1) at (0, -1.5) {\contour{white}{$s_{1}$}};
    \node[draw, circle, q2class] (q20) at (-0.5, -3.0) {\contour{white}{$s_{20}$}};
    \node[draw, circle, q2class] (q21) at (0.5, -3.0) {\contour{white}{$s_{21}$}};
    \node[left=0.3cm of q00] (start) {};

    \path[->,>=stealth',every node/.style={font=\sffamily\scriptsize}]
        (q00) edge[] node[left, align=center] {$a/A$\\ $b/B$} (q1)
        (q01) edge[] node[right, align=center] {$a/A$\\ $b/B$} (q1)
        (q1) edge[] node[left, align=center] {$a/B$} (q20)
        (q1) edge[] node[right, align=center] {$b/C$} (q21)
        (q20) edge[bend left=40] node[right, align=center] {$a/C$\\ $b/A$} (q00)
        (q21) edge[bend right=40] node[left, align=center] {$a/C$\\ $b/A$} (q01)
        (start) edge[] (q00)
        ;
\end{tikzpicture}
        \caption{$2$-step confluent expansion $\mathcal{M}_1$}
        \label{fig:valid-expansion-m1}
    \end{subfigure}
    \begin{subfigure}{0.59\linewidth}
        \centering
        \begin{tikzpicture}
    \node[draw, circle, q0class] (q00) at (-0.65, 0) {\contour{white}{$s_{00}$}};
    \node[draw, circle, q0class] (q01) at (0.65, 0) {\contour{white}{$s_{01}$}};
    \node[draw, circle, q1class] (q10) at (-0.65, -1.5) {\contour{white}{$s_{10}$}};
    \node[draw, circle, q1class] (q11) at (0.65, -1.5) {\contour{white}{$s_{11}$}};
    \node[draw, circle, q2class] (q20) at (-0.65, -3.0) {\contour{white}{$s_{20}$}};
    \node[draw, circle, q2class] (q21) at (0.65, -3.0) {\contour{white}{$s_{21}$}};
    \node[left=0.3cm of q00] (start) {};

    \path[->,>=stealth',every node/.style={font=\sffamily\scriptsize}]
        (q00) edge[] node[below, sloped, align=center] {$a/A$} (q10)
        (q00) edge[bend left=2] node[above, sloped, xshift=6pt, align=center] {$b/B$} (q11)
        (q01) edge[bend right=2] node[above, sloped, xshift=-6pt, align=center] {$b/B$} (q10)
        (q01) edge[] node[above, sloped, align=center] {$a/A$} (q11)
        (q10) edge[] node[left, align=center] {$a/B$\\ $b/C$} (q20)
        (q11) edge[] node[right, align=center] {$a/B$\\ $b/C$} (q21)
        (q20) edge[bend left=70] node[right, align=center] {$a/C$\\ $b/A$} (q00)
        (q21) edge[bend right=70] node[left, align=center] {$a/C$\\ $b/A$} (q01)
        (start) edge[] (q00)
        ;
\end{tikzpicture}\hspace{-0.2cm}\begin{tikzpicture}
    \node[draw, circle, q0class] (q00) at (-0.75, 0) {\contour{white}{$s_{00}$}};
    \node[draw, circle, q0class] (q01) at (0.75, -3.0) {\contour{white}{$s_{01}$}};
    \node[draw, circle, q1class] (q10) at (-0.75, -1.5) {\contour{white}{$s_{10}$}};
    \node[draw, circle, q1class] (q11) at (0.75, -1.5) {\contour{white}{$s_{11}$}};
    \node[draw, circle, q2class] (q20) at (-0.75, -3.0) {\contour{white}{$s_{20}$}};
    \node[draw, circle, q2class] (q21) at (0.75, 0) {\contour{white}{$s_{21}$}};
    \node[left=0.3cm of q00] (start) {};

    \path[->,>=stealth',every node/.style={font=\sffamily\scriptsize}]
        (q00) edge[] node[below, sloped, align=center] {$a/A$\\ $b/B$} (q10)
        (q10) edge[] node[below, sloped, align=center] {$a/B$\\ $b/C$} (q20)
        (q20) edge[] node[above, sloped, align=center] {$a/C$\\ $b/A$} (q01)
        (q01) edge[] node[below, sloped, align=center] {$a/A$\\ $b/B$} (q11)
        (q11) edge[] node[below, sloped, align=center] {$a/B$\\ $b/C$} (q21)
        (q21) edge[] node[below, sloped, align=center] {$a/C$\\ $b/A$} (q00)
        (start) edge[] (q00)
        ;
\end{tikzpicture}
        \caption{Two non-confluent expansions}
        \label{fig:invalid-expansion-m1}
    \end{subfigure}
    \caption{Confluent and non-confluent expansions of $\mathcal{M'}$}
    \label{fig:example1}
\end{figure}

\begin{example}
    \label{ex:example1}
    Consider the minimal \ac{MM} $\mathcal{M}'$ illustrated in \Cref{fig:example1_min} with inputs $a$ and $b$.
    \Cref{fig:example1} shows three bisimilar \ac{MM} expansions of $\mathcal{M}'$.
    $\mathcal{M}_1$ is an expansion satisfying the \emph{2-step confluence} criterion as it is indeed guaranteed that from every pair of \ac{I/O} equivalent states (e.g., $s_{20}$ and $s_{21}$) the same state $s_1$ will be reached after at most $d = 2$ steps.
    This does not apply to the other two examples, as in both cases from $s_{00}$ and $s_{01}$ the same node will never be reached when repeatedly applying the input sequence $aaa$.
\end{example}

We might impose additional constraints to further reduce the number of possible 
expansions.
An instrumental constraint in this regard stipulates that the root state of the minimal \ac{MM} has not been expanded:
\begin{definition}
    \label{def:unique-root}
    An \ac{MDM} $\mathcal{D} = (\splitatcommas{S, I, O, s_{0}, \delta, \lambda, D})$ has a unique root iff there is no state $s\in Q$ different from $s_0$ such that $s_0 \io s$.
\end{definition}
Another constraint disallows variations regarding delay behaviors across sink nodes, echoing that in a sink node the future behavior is considered not of interest any longer, something which we will assume across all empirical studies:
\begin{definition}
    \label{def:sink-delay-ignorant}
    We denote an \ac{MDM} $\mathcal{D} = (S, I, O, s_{0}, \delta, \lambda, D)$ as sink delay-ignorant iff for each state $s\in S$, where $\delta(s, i) \io s$ for each input $i \in I$, it holds $\delta(s, i) = s$.
\end{definition}

\subsection{Sampling-Based $L^*$ Learning}
\label{sec:extended-sampling-aal}
As a starting point, we modify the classical $L^*$ algorithm such that it additionally separates states based on their delay behavior according to the $d$-step confluence criterion.
In this criterion, we know that any two \ac{I/O}-equivalent states exhibit distinguishable delay behavior for $d$ steps at most. This behavior can be tested by test words with a maximum length of $d$. Thus, the easiest way to integrate the state separation into $L^*$ is by initializing the test set $T$ by all non-empty words $I^{\leq d}$ and sampling all queries $(Q \cup \{qi \mid q\in Q, i\in I\}) \times T$ $k$~times.
Otherwise, we can continue the general learning procedure of $L^*$ as usual by adding any test word longer than $d$ that arises by a counterexample (where, however, each newly added test word only needs to be queried once).

Obviously, the size of the test set $T$ now has an exponential size regarding the parameter $d$, and one could argue that it is better to gradually increase $T$ through counterexamples. However, we consider \acp{EQ} to be unrealistic to pose regarding delay behavior. One possibility to implement an \ac{EO} nonetheless would be by adapting the $W$-method~\cite{W-method, W-method-2} while repeating each query $k$ times, but the size of the test suite would still grow exponentially by the maximum number of additional states of the true \ac{SUL}. In the context of $d$-step confluence, we thus would not gain significant advantages by offloading this part to the \ac{EO}.

\subsection{Efficient Coverage Based on Mealy Machine Expansion}
\label{sec:extended-efficient-coverage-mealy}

We now present how to leverage the approach in~\Cref{sec:efficient-coverage-mealy} for a procedure addressing the more general case. We perform the following steps:
\begin{enumerate}
    \item We start by inferring the underlying minimal \ac{MM} $\mathcal{M}'$ of the \ac{SUL} by regular \ac{AAL} algorithms, e.g., $L^*$.
    \item Based on the assumptions on the stutter-free $d$-step confluence of the \ac{SUL}~$\mathcal{D}$, we expand $\mathcal{M}'$ into an \ac{MM} $\mathcal{M}^d$, which we use to sample delays.
    \item We collect sampling information for $\mathcal{M}^d$ as described in \Cref{sec:efficient-coverage-mealy}, thereby arriving at $\mathcal{D}^d$.
    \item Lastly, we check for delay equivalence of states in $\mathcal{D}^d$ and merge those states that have statistically equivalent delay behavior to obtain     $\mathcal{D}'$.
\end{enumerate}
In the following, the second and fourth steps will be explained in detail.

\paragraph{Expansion of learned \ac{MM}.}

Based on the learned minimal \ac{MM} $\mathcal{M}'$, we now create an expanded \ac{MM}, which we use as basis to collect delay information, under the assumption that the SUL \ac{MDM} $\mathcal{D}$ satisfies the stutter-free $d$-step confluence criterion.
We call this machine in the following $\mathcal{M}^d$. For the construction of $\mathcal{M}^d$, we need to ensure that, when two states in the actual system are different, this also needs to be the case for $\mathcal{M}^d$. We call this property \emph{injective bisimilarity}:
\begin{definition}
    \label{def:refining-bisimulation}
    We say that \ac{MM} $\mathcal{M} = (S, I, O, s_{0}, \delta, \lambda)$ is injective bisimilar to \ac{MM} $\mathcal{M}' = (S', I, O, s'_{0}, \delta', \lambda')$ iff there exists a bisimulation relation between $S$ and $S'$ that is injective.
\end{definition}

Under the assumption that every conceivable \ac{SUL} $\mathcal{D}$ satisfies the criterion of \Cref{def:no-self-loops} and \Cref{def:d-step-confluence}, we can construct an \ac{MM} $\mathcal{M}^d$ from the learned minimal \ac{MM} $\mathcal{M}' = (S', I, O, s'_0, \delta', \lambda')$ as follows: For each state $s' \in S'$, we consider how $s'$ could be reached from any other state in $d$ steps, or from the root $s'_0$ in less than $d$ steps.
More formally, we consider all pairs $\langle p', w\rangle$ with states $p' \in S'$ and loop-free input words $w \in I^*$ such that $\delta'(p', w) = s'$ with $|w| = d$ or $\delta'(s'_0, w) = s'$ with $|w| < d$.
In this context, we say that a word $w\in I^*$ starting from $p\in S$ is \emph{loop-free} iff $w$ does not stutter on any self-loop, i.e., $\delta(p, w_1...w_j) \notio \delta(p, w_1...w_{j + 1})$ for each $1 \leq j < |w|$.
For each pair $\langle p', w\rangle$, we create one state for the resulting \ac{MM} $\mathcal{M}^d$, which we shall refer to  as $s^d_{s', \langle p', w\rangle}$.

We add a transition triggered by input $i \in I$ between the two states $s^d_{s', \langle p', w\rangle}$ and $t^d_{t', \langle r', l\rangle}$ if there is a transition triggered by $i$ from $s'$ to $t'$ and both states are compatible regarding their transition history.
This is the case when either $\delta'(p', w_1) = r'$ and $w_2...w_{|w|}i = l$ (for $|w| = d$) or $p' = r' = s'_0$ and $wi = l$ (for $|w| < d$).
For any self-loop $\delta'(s', i) = s'$ in the minimal \ac{MM} $\mathcal{M}'$, we also create a self-loop in $\mathcal{M}^d$ with $\delta^d(s^d_{s', \langle p'
, w\rangle}, i) = s^d_{s', \langle p'
, w\rangle}$.
We set the output to be equivalent to the minimal \ac{MM}: $\lambda^d(s^d_{s', \langle p', w\rangle}, i) = \lambda'(s', i)$.

We can show that this construction of $\mathcal{M}^d$ has the required properties:
\begin{theorem}
    \label{thm:Mstar-inj-bisim}
    Let $\mathcal{D}$ be a stutter-free $d$-step confluent \ac{MDM} with underlying \ac{MM} $\mathcal{M}$. Then $\mathcal{M}$ is injective bisimilar to the \ac{MM} $\mathcal{M}^d$ constructed as proposed above from the minimal \ac{MM} $\mathcal{M}'$ of $\mathcal{M}$.
\end{theorem}

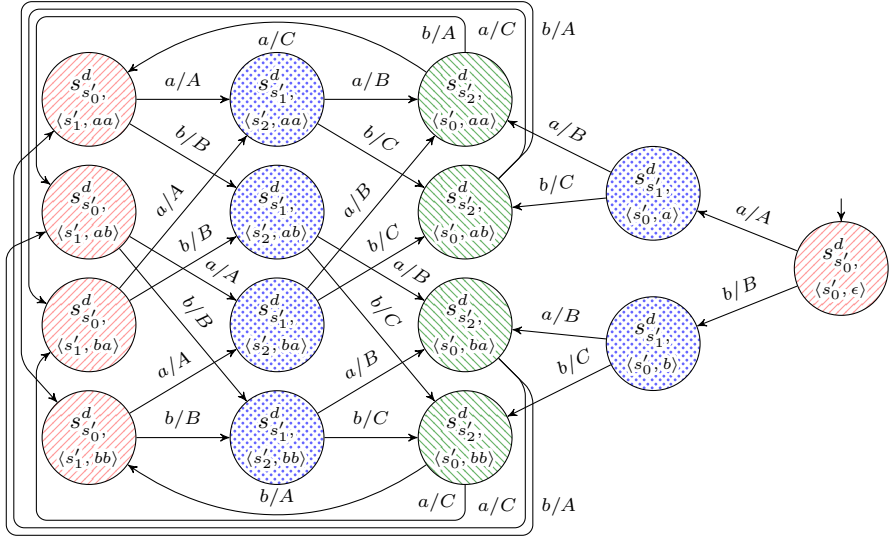
\begin{figure}[t]
    \begin{tikzpicture}
    \node[draw, circle, align=center, inner sep=0pt, text width=0.95cm, q0class] (q0aa) at (0, 0) {\shortstack{\contour{white}{$s^d_{s'_0,}$}\\ \contour{white}{\textsubscript{$\langle s'_1, aa\rangle$}}}};
    \node[draw, circle, align=center, inner sep=0pt, text width=0.95cm, q0class] (q0ab) at (0, -1.5) {\shortstack{\contour{white}{$s^d_{s'_0,}$}\\ \contour{white}{\textsubscript{$\langle s'_1, ab\rangle$}}}};
    \node[draw, circle, align=center, inner sep=0pt, text width=0.95cm, q0class] (q0ba) at (0, -3) {\shortstack{\contour{white}{$s^d_{s'_0,}$}\\ \contour{white}{\textsubscript{$\langle s'_1, ba\rangle$}}}};
    \node[draw, circle, align=center, inner sep=0pt, text width=0.95cm, q0class] (q0bb) at (0, -4.5) {\shortstack{\contour{white}{$s^d_{s'_0,}$}\\ \contour{white}{\textsubscript{$\langle s'_1, bb\rangle$}}}};

    \node[draw, circle, align=center, inner sep=0pt, text width=0.95cm, q1class] (q1aa) at (2.5, 0) {\shortstack{\contour{white}{$s^d_{s'_1,}$}\\ \contour{white}{\textsubscript{$\langle s'_2, aa\rangle$}}}};
    \node[draw, circle, align=center, inner sep=0pt, text width=0.95cm, q1class] (q1ab) at (2.5, -1.5) {\shortstack{\contour{white}{$s^d_{s'_1,}$}\\ \contour{white}{\textsubscript{$\langle s'_2, ab\rangle$}}}};
    \node[draw, circle, align=center, inner sep=0pt, text width=0.95cm, q1class] (q1ba) at (2.5, -3) {\shortstack{\contour{white}{$s^d_{s'_1,}$}\\ \contour{white}{\textsubscript{$\langle s'_2, ba\rangle$}}}};
    \node[draw, circle, align=center, inner sep=0pt, text width=0.95cm, q1class] (q1bb) at (2.5, -4.5) {\shortstack{\contour{white}{$s^d_{s'_1,}$}\\ \contour{white}{\textsubscript{$\langle s'_2, bb\rangle$}}}};

    \node[draw, circle, align=center, inner sep=0pt, text width=0.95cm, q2class] (q2aa) at (5, 0) {\shortstack{\contour{white}{$s^d_{s'_2,}$}\\ \contour{white}{\textsubscript{$\langle s'_0, aa\rangle$}}}};
    \node[draw, circle, align=center, inner sep=0pt, text width=0.95cm, q2class] (q2ab) at (5, -1.5) {\shortstack{\contour{white}{$s^d_{s'_2,}$}\\ \contour{white}{\textsubscript{$\langle s'_0, ab\rangle$}}}};
    \node[draw, circle, align=center, inner sep=0pt, text width=0.95cm, q2class] (q2ba) at (5, -3) {\shortstack{\contour{white}{$s^d_{s'_2,}$}\\ \contour{white}{\textsubscript{$\langle s'_0, ba\rangle$}}}};
    \node[draw, circle, align=center, inner sep=0pt, text width=0.95cm, q2class] (q2bb) at (5, -4.5) {\shortstack{\contour{white}{$s^d_{s'_2,}$}\\ \contour{white}{\textsubscript{$\langle s'_0, bb\rangle$}}}};

    \node[draw, circle, align=center, inner sep=0pt, text width=0.95cm, q0class] (q0) at (10, -2.25) {\shortstack{\contour{white}{$s^d_{s'_0,}$}\\ \contour{white}{\textsubscript{$\langle s'_0, \epsilon\rangle$}}}};
    \node[draw, circle, align=center, inner sep=0pt, text width=0.95cm, q1class] (q1a) at (7.5, -1.25) {\shortstack{\contour{white}{$s^d_{s'_1,}$}\\ \contour{white}{\textsubscript{$\langle s'_0, a\rangle$}}}};
    \node[draw, circle, align=center, inner sep=0pt, text width=0.95cm, q1class] (q1b) at (7.5, -3.25) {\shortstack{\contour{white}{$s^d_{s'_1,}$}\\ \contour{white}{\textsubscript{$\langle s'_0, b\rangle$}}}};

    \node[above=0.3cm of q0] (start) {};

    \path[->,>=stealth',every node/.style={font=\sffamily\scriptsize}]
        (q0aa) edge[] node[above, sloped, align=center] {$a/A$} (q1aa)
               edge[] node[above, sloped, align=center] {$b/B$} (q1ab)
        (q0ab) edge[] node[above right, xshift=3pt, sloped, align=center] {$a/A$} (q1ba)
               edge[] node[above, sloped, align=center] {$b/B$} (q1bb)
        (q0ba) edge[] node[above, sloped, align=center] {$a/A$} (q1aa)
               edge[] node[above right, sloped, align=center] {$b/B$} (q1ab)
        (q0bb) edge[] node[above, sloped, align=center] {$a/A$} (q1ba)
               edge[] node[above, sloped, align=center] {$b/B$} (q1bb)

        (q1aa) edge[] node[above, sloped, align=center] {$a/B$} (q2aa)
               edge[] node[above, sloped, align=center] {$b/C$} (q2ab)
        (q1ab) edge[] node[above right, xshift=3pt, sloped, align=center] {$a/B$} (q2ba)
               edge[] node[above, sloped, align=center] {$b/C$} (q2bb)
        (q1ba) edge[] node[above, sloped, align=center] {$a/B$} (q2aa)
               edge[] node[above right, sloped, align=center] {$b/C$} (q2ab)
        (q1bb) edge[] node[above, sloped, align=center] {$a/B$} (q2ba)
               edge[] node[above, sloped, align=center] {$b/C$} (q2bb)

        (q2aa) edge[bend right=35] node[below, sloped, align=center] {$a/C$} (q0aa)
        (q2bb) edge[bend left=35] node[above, sloped, align=center] {$b/A$} (q0bb)

        (q0)   edge[] node[above, sloped, align=center] {$a/A$} (q1a)
               edge[] node[above, sloped, align=center] {$b/B$} (q1b)
        (q1a)  edge[] node[above, sloped, align=center] {$a/B$} (q2aa)
               edge[] node[above, sloped, align=center] {$b/C$} (q2ab)
        (q1b)  edge[] node[above, sloped, align=center] {$a/B$} (q2ba)
               edge[] node[above right, sloped, align=center] {$b/C$} (q2bb)

        (start) edge[] (q0)
        ;

    \path[draw,rounded corners,->,>=stealth',every node/.style={font=\sffamily\scriptsize}]
        (q2aa) -- node[left] {$b/A$} ++(0, 1.1) -- ++(-5.7, 0) -- ++(0, -2.1) -- (q0ab);

    \path[draw,rounded corners,->,>=stealth',every node/.style={font=\sffamily\scriptsize}]
        (q2ab) -- ++(0.8, 0.8) -- node[left, pos=0.85] {$a/C$} ++(0, 1.9) -- ++(-6.6, 0) -- ++(0, -3.8) -- (q0ba);
    \path[draw,rounded corners,->,>=stealth',every node/.style={font=\sffamily\scriptsize}]
        (q2ab) -- ++(0.9, 0.9) -- node[right, pos=0.8] {$b/A$} ++(0, 1.9) -- ++(-6.8, 0) -- ++(0, -4.8) -- (q0bb);

    \path[draw,rounded corners,->,>=stealth',every node/.style={font=\sffamily\scriptsize}]
        (q2ba) -- ++(0.8, -0.8) -- node[left, pos=0.85] {$a/C$} ++(0, -1.9) -- ++(-6.8, 0) -- ++(0, 4.8) -- (q0aa);
    \path[draw,rounded corners,->,>=stealth',every node/.style={font=\sffamily\scriptsize}]
        (q2ba) -- ++(0.9, -0.9) -- node[right, pos=0.8] {$b/A$} ++(0, -1.9) -- ++(-7.0, 0) -- ++(0, 3.8) -- (q0ab);

    \path[draw,rounded corners,->,>=stealth',every node/.style={font=\sffamily\scriptsize}]
        (q2bb) -- node[left] {$a/C$} ++(0, -1.1) -- ++(-5.7, 0) -- ++(0, +2.1) -- (q0ba);
\end{tikzpicture}
    \caption{Expanded $\mathcal{M}^d$ for minimal \ac{MM} $\mathcal{M}'$ from \Cref{fig:example1_min} for depth $d = 2$}
    \label{fig:example-unfolding}
\end{figure}
\begin{example}
    Let us recall the \ac{MM} $\mathcal{M}'$ from \Cref{ex:example1}. The expanded \ac{MM} $\mathcal{M}^d$  for $d = 2$ is presented in \Cref{fig:example-unfolding}. As we have two different inputs and are tracking two steps back, we start by creating $2\cdot 2 = 4$ states for each state from the original automaton. In the end, we need to add three states to identify the correct state when starting from the root.
    It can be easily verified that the \ac{MM} $\mathcal{M}_1$ from \Cref{fig:valid-expansion-m1} is injective bisimilar to $\mathcal{M}^d$.
    On the other hand, when only constructing $\mathcal{M}^d$ for $d = 1$, the injective bisimulation property is no longer fulfilled for $\mathcal{M}_1$, as for identifying the state $s_{00}$ and $s_{01}$ by the transition history beginning with $s_1$, we need a horizon of at least two steps.
\end{example}

The resulting construction is linear in the size of the state space $S'$ of the minimal \ac{MM} $\mathcal{M}'$ and exponential only in the selection of $d$:
\begin{theorem}
    \label{thm:size-expanded-machine}
    $|\mathcal{M}^d| \in \mathcal{O}(|S'| \cdot |I|^d)$.
\end{theorem}

Some further restrictions can easily be applied: When assuming a unique root state (\Cref{def:unique-root}), the additional construction to identify the proper state after $d$ steps can be omitted, and no state splitting is necessary. Similarly, when assuming a sink delay-ignorant \ac{MDM} (\Cref{def:sink-delay-ignorant}), no splitting of sink states is required.

\myparagraph{Selection of $d$.} In general, it is unknown how large the selected $d$ should be, but this is a general problem across \ac{AAL} (shared for instance with the $W$-method~\cite{W-method, W-method-2} where the maximum additional number of states $e$ is notoriously unknown). Furthermore, there are examples where no suitable $d$ can be found at all, see~\Cref{fig:invalid-expansion-m1}.
In this case, the bisimulation relation between $\mathcal{M}$ and $\mathcal{M}^d$ is not injective, and for one state in $\mathcal{M}^d$, there can be multiple states corresponding to $\mathcal{M}$. Then, we would obtain an uncontrolled mixture of multiple \acp{CDF} (depending on how often we actually visit each corresponding transition in $\mathcal{M}$).
Nevertheless, increasing $d$ never worsens the result, as a consequence of the following lemma:
\begin{lemma}
    \label{thm:increasing-d-smaller}
    For each $d\in \mathbb{N}_0$, $\mathcal{M}^d$ is injective bisimilar to $\mathcal{M}^{d+1}$.
\end{lemma}
So, if two states from $\mathcal{M}$ are correctly distinguished in $\mathcal{M}^d$, then also in $\mathcal{M}^{d+1}$.

\myparagraph{Other restrictions.} Note that the concept of injective bisimulations is not restricted to stutter-free $d$-step confluenct \acp{MDM}. By essentially using the cross-product of the \acp{MM}, as long as the number of actual state structures given a minimal \ac{MM} $\mathcal{M}'$ is finite, a suitable \ac{MM} $\mathcal{M}^*$ can be constructed as follows: 
\begin{theorem}
    \label{thm:feasible-inj-bisim}
    Given pairwise bisimilar \acp{MM} $\mathcal{M}_1, ..., \mathcal{M}_n$, one can find an \ac{MM} $\mathcal{M}^*$ with, at maximum, $|\mathcal{M}_1| \cdot ...\cdot |\mathcal{M}_n|$ states such that every $\mathcal{M}_j$ with $j\in \{1, ..., n\}$ is injective bisimilar to $\mathcal{M}^*$.
\end{theorem}
In particular, we can use this theorem on an enumeration of all \acp{MM} that are bisimilar to the learned minimal representation and at most as large as some given size~$b$.
However, the construction leads to a vast number of states in $\mathcal{M}^*$. Thus, the challenge for other restrictions as proposed in this paper lies in finding a reasonable balance between expressiveness and size of the expanded \ac{MM}.

\paragraph{State merging.} The merging step can be considered optional. It is meant to ensure the minimality of the resulting \ac{MDM}.
To perform the merging step, we need to check for each pair of states $s$ and $t$ if it holds $s \ido t$. If $s \notio t$, it trivially also follows $s \notido t$. Otherwise, we have to resort to a test that checks if the delay distributions are equal~\cite{rubner:1998:earth-mover,two-sample-ks-test,compare-distributions-2} as discussed in \Cref{sec:background}.
Should the transitivity property of $\ido$ not be justifiable statistically, the following options arise:
\begin{itemize}
    \item Decide not to merge two state clusters if two distributions in the merged cluster are not equal according to the test for distribution equality.
    \item Apply a clustering method \cite{nielsen:2014:alpha-divergence, henderson:2015:ep-means} from \Cref{sec:background}.
    \item Acquire more samples\footnote{This takes up a remark from \Cref{sec:delay-mealy-machine} that our method supports to work with different $k$ for some sub-structures, here effectuated  incrementally in another \ac{RPP} construction.} for disambiguation of selected transitions, and retry the merging procedure subsequently.
\end{itemize}
As the construction using the expanded \ac{MM} determines the size of $\mathcal{M}^d$, the sampling effort is independent of the true \ac{SUL}'s delay behavior. This also applies to an unstable (e.g., unreliably returning wrong outcomes) \ac{CDF} equivalence test.
On the other hand, an unstable test in sampling-based $L^*$ learning can lead to a runtime explosion, when equal distributions are repeatedly reported as different.

\section{Empirical Evaluation}
\label{sec:evaluation}

In this section, we evaluate the efficacy of both approaches, the baseline implementation based on the $L^*$ algorithm (\Cref{sec:sampling-aal} and \Cref{sec:extended-sampling-aal}) as well as the improved method with additional collection of samples (\Cref{sec:efficient-coverage-mealy} and \Cref{sec:extended-efficient-coverage-mealy}).
For the second approach, we also use $L^*$ to learn the \ac{I/O} behavior. Although we could have resorted to better-performing alternatives, e.g., TTT~\cite{TTT} or $L^\#$~\cite{Lsharp-algorithm}, the impact on the final runtime is negligible as the majority of steps are needed for the subsequent sampling of delays.
Furthermore, we do not require that self-loops are sampled $k$ times in the expanded \ac{MM}, given the assumed stutter-freeness (see \Cref{def:no-self-loops}).
Both approaches were implemented on top of AALpy~\cite{AALpy}.
The research questions in focus of our benchmarking studies are the following:

\begin{description}
    \item[RQ1:] How is the performance of both approaches compared to each other?
    \item[RQ2:] What is the impact of $d$ in the $d$-step confluence criterion on the runtime?
    \item[RQ3:] How does the alphabet size $|I|$ influence the methods' scalability?
    \item[RQ4:] What changes if making the assumption of a unique root (\Cref{def:unique-root})?
\end{description}

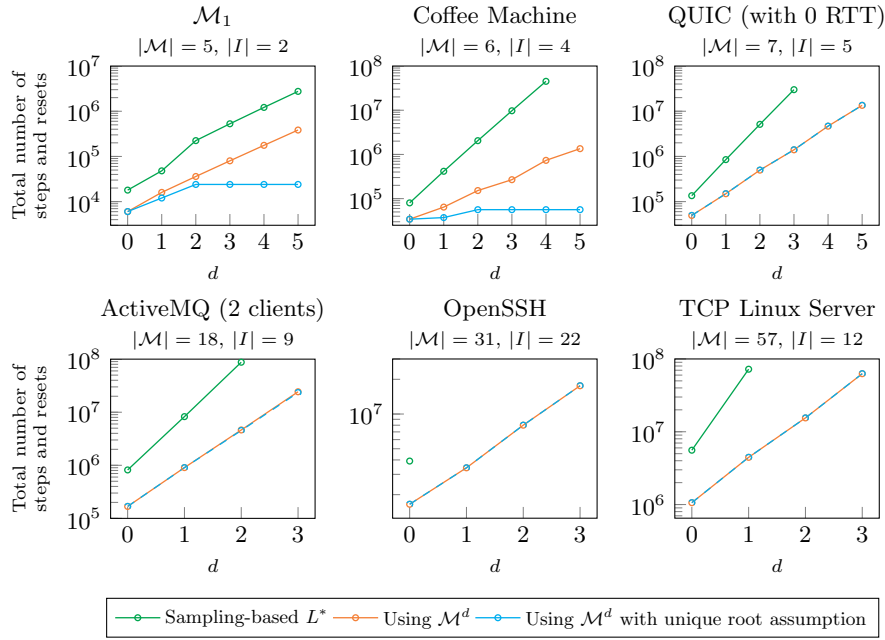
\begin{figure}[t]
    \centering
    \begin{tikzpicture}

\begin{axis}[
    at={(0cm, 0cm)}, anchor=south east,
    legend style={nodes={scale=0.5, transform shape}},
    title style={yshift=-1.5ex,align=center},
    title={$\mathcal{M}_1$ \\ \scriptsize{$|\mathcal{M}| = 5$, $|I| = 2$}},
    width=4.3cm,height=3.7cm,
    ymode=log,
    xlabel=\scriptsize{$d$},
    ylabel style={align=center},
    ylabel={\shortstack{\scriptsize{Total number of} \\ \scriptsize{steps and resets}}},
    xtick={0, 1, ..., 6},
    ytick={100, 1000, 10000, 100000, 1000000, 10000000, 100000000},
    xtick pos=bottom,
    ytick pos=left,
    ymin=3000, ymax=10000000,
    name=plot1,
]

\addplot[
    color=Green,
    mark=o,
    mark size=1.1pt,
    line width=0.5pt
    ]
    coordinates {
(0, 17998)(1, 48000)(2, 224000)(3, 528000)(4, 1216000)(5, 2752000)
};

\addplot[
    color=Orange,
    mark=o,
    mark size=1.1pt,
    line width=0.5pt
    ]
    coordinates {(0, 6025)(1, 16013)(2, 36013)(3, 80017)(4, 176029)(5, 384049)
};

\addplot[
    color=Cyan,
    mark=o,
    mark size=1.1pt,
    line width=0.5pt
    ]
    coordinates {(0, 6025)(1, 12025)(2, 24025)(3, 24025)(4, 24025)(5, 24025)
};

\end{axis}

\begin{axis}[
    at={(3.75cm, 0cm)}, anchor=south east,
    legend style={nodes={scale=0.5, transform shape}},
    title style={yshift=-1.5ex,align=center},
    title={Coffee Machine\\ \scriptsize{$|\mathcal{M}| = 6$, $|I| = 4$}},
    width=4.3cm,height=3.7cm,
    ymode=log,
    xlabel=\scriptsize{$d$},
    xtick={0, 1, 2, 3, 4, 5},
    ytick={1000, 10000, 100000, 1000000, 10000000, 100000000},
    xtick pos=bottom,
    ytick pos=left,
    ymin=25000, ymax=100000000,
    legend pos=north west,
    name=plot3,
    ]

    \addplot[
        color=Green,
        mark=o,
        mark size=1.1pt,
        line width=0.5pt
    ]
    coordinates {
(0, 80162)(1, 416000)(2, 2048000)(3, 9728000)(4, 45056000)
};

    \addplot[
        color=Orange,
        mark=o,
        mark size=1.1pt,
        line width=0.5pt
    ]
    coordinates {
(0, 34282)(1, 64317)(2, 152339)(3, 268339)(4, 734339)(5, 1350339)
};

    \addplot[
        color=Cyan,
        mark=o,
        mark size=1.1pt,
        line width=0.5pt
    ]
    coordinates {
(0, 34282)(1, 37282)(2, 56282)(3, 56282)(4, 56282)(5, 56282)
    };

\end{axis}

\begin{axis}[
    at={(7.5cm, 0cm)}, anchor=south east,
    legend style={nodes={scale=0.5, transform shape}},
    title style={yshift=-1.5ex,align=center},
    title={QUIC (with $0$ RTT)\\ \scriptsize{$|\mathcal{M}| = 7$, $|I| = 5$}},
    width=4.3cm,height=3.7cm,
    ymode=log,
    xlabel=\scriptsize{$d$},
    xtick={0, 1, 2, 3, 4, 5},
    ytick={1000, 10000, 100000, 1000000, 10000000, 100000000},
    xtick pos=bottom,
    ytick pos=left,
    ymin=30000, ymax=100000000,
    legend pos=north west,
    name=plot3,
    ]

    \addplot[
        color=Green,
        mark=o,
        mark size=1.1pt,
        line width=0.5pt
    ]
    coordinates {
(0, 135356)(1, 850000)(2, 5125000)(3, 30000000)
    };

    \addplot[
        color=Orange,
        mark=o,
        mark size=1.1pt,
        line width=0.5pt,
    ]
    coordinates {
(0, 49433)(1, 149628)(2, 497839)(3, 1401020)(4, 4674142)(5, 13493213)
    };

    \addplot[
        color=Cyan,
        mark=o,
        mark size=1.1pt,
        line width=0.5pt,
        dashed
    ]
    coordinates {
(0, 49433)(1, 149628)(2, 497839)(3, 1401020)(4, 4674142)(5, 13493213)
    };

\end{axis}

\begin{axis}[
    at={(0cm, -3.9cm)}, anchor=south east,
    legend style={nodes={scale=0.5, transform shape}},
    title style={yshift=-1.5ex,align=center},
    title={ActiveMQ (2 clients)\\ \scriptsize{$|\mathcal{M}| = 18$, $|I| = 9$}},
    width=4.3cm,height=3.7cm,
    ymode=log,
    xlabel=\scriptsize{$d$},
    xtick={0, 1, 2, 3},
    ylabel style={align=center},
    ylabel={\shortstack{\scriptsize{Total number of} \\ \scriptsize{steps and resets}}},
    ytick={10000, 100000, 1000000, 10000000, 100000000},
    xtick pos=bottom,
    ytick pos=left,
    ymin=100000, ymax=100000000,
    legend pos=north west,
    name=plot4,
    ]

    \addplot[
        color=Green,
        mark=o,
        mark size=1.1pt,
        line width=0.5pt
    ]
    coordinates {
(0, 817461)(1, 8262000)(2, 87480000)
    };

    \addplot[
        color=Orange,
        mark=o,
        mark size=1.1pt,
        line width=0.5pt
        ]
    coordinates {
(0, 167895)(1, 904764)(2, 4628528)(3, 24285403)
        };

    \addplot[
        color=Cyan,
        mark=o,
        mark size=1.1pt,
        line width=0.5pt,
        dashed
        ]
        coordinates {
(0, 167895)(1, 903125)(2, 4525555)(3, 23438659)
        };

\end{axis}

\begin{axis}[
    at={(3.75cm, -3.9cm)}, anchor=south east,
    legend style={nodes={scale=0.5, transform shape}},
    title style={yshift=-1.5ex,align=center},
    title={OpenSSH\\ \scriptsize{$|\mathcal{M}| = 31$, $|I| = 22$}},
    width=4.3cm,height=3.7cm,
    ymode=log,
    xlabel=\scriptsize{$d$},
    xtick={0, 1, 2, 3},
    ytick={10000, 100000, 1000000, 10000000, 100000000, 1000000000},
    xtick pos=bottom,
    ytick pos=left,
    ymin=1250000, ymax=30000000,
    legend pos=north west,
    name=plot5,
    ]

    \addplot[
        color=Green,
        mark=o,
        mark size=1.1pt,
        line width=0.5pt
    ]
    coordinates {
(0, 3911647)
    };

    \addplot[
        color=Orange,
        mark=o,
        mark size=1.1pt,
        line width=0.5pt
        ]
        coordinates {
(0, 1651532)(1, 3418520)(2, 8003874)(3, 17582305)
        };

    \addplot[
        color=Cyan,
        mark=o,
        mark size=1.1pt,
        line width=0.5pt,
        dashed
        ]
        coordinates {
(0, 1651532)(1, 3418520)(2, 8003874)(3, 17582305)
        };

\end{axis}

\begin{axis}[
    at={(7.5cm, -3.9cm)}, anchor=south east,
    legend style={legend columns=-1, nodes={scale=0.8, transform shape}, at={(1.0, -0.5)}},
    legend image post style={solid},
    title style={yshift=-1.5ex,align=center},
    title={TCP Linux Server\\ \scriptsize{$|\mathcal{M}| = 57$, $|I| = 12$}},
    width=4.3cm,height=3.7cm,
    ymode=log,
    xlabel=\scriptsize{$d$},
    xtick={0, 1, 2, 3},
    ytick={10000, 100000, 1000000, 10000000, 100000000},
    xtick pos=bottom,
    ytick pos=left,
    ymin=650000, ymax=100000000,
    name=plot6,
    ]

    \addplot[
    color=Green,
    mark=o,
    mark size=1.1pt,
        line width=0.5pt
    ]
    coordinates {
(0, 5579750)(1, 72288000)
    };
    \addlegendentry{Sampling-based $L^*$}

    \addplot[
        color=Orange,
        mark=o,
        mark size=1.1pt,
        line width=0.5pt
        ]
        coordinates {
(0, 1060854)(1, 4451521)(2, 15494497)(3, 62752520)
        };
    \addlegendentry{Using $\mathcal{M}^d$}

    \addplot[
    color=Cyan,
    mark=o,
    mark size=1.1pt,
    line width=0.5pt,
    dashed
    ]
    coordinates {
(0, 1060854)(1, 4451521)(2, 15494497)(3, 62752520)
    };
    \addlegendentry{Using $\mathcal{M}^d$ with unique root assumption}
\end{axis}
\end{tikzpicture}
    \caption{Number of total actions for different models and selections of $d$}
    \label{fig:experiment-1}
\end{figure}
\myparagraph{Experimental setup.}
We considered the \ac{MM} $\mathcal{M}_1$ from \Cref{fig:valid-expansion-m1} together with cases from a publicly available test suite for \acp{MM}~\cite{automata-test-suite}, a selection of which is presented here.
Even though the \acp{MM} from \cite{automata-test-suite} are minimal regarding their \ac{I/O} behavior, they can be used to assess the impact of parameter~$d$, corresponding to an assumed need for state separation caused by the delay behavior.
Ideally, the state merging step leads to a final merge back to the \ac{I/O} minimal result. 
The delay of each transition is synthetic, based on exponential distributions with random rates.
We assume two distributions to be different if the relative deviation of the sample mean is more than $20$ percent and the absolute deviation is, at least, $0.01$. Each transition has to be visited at least $k = 1000$ times (for parameter selection see \Cref{sec:bounding-relative-error-exponential}). As the number of performed actions per experiment can grow large rapidly, executions exceeding $10^8$ actions are aborted.

\myparagraph{Measurement data.} In \Cref{fig:experiment-1}, the results for a selection of experiments are depicted, showing the number of total steps and resets depending on the selection of the parameter $d$. While the upper row presents studies of small examples, the lower row covers larger real-world examples. 
All results from the test suite~\cite{automata-test-suite} can be found in~\Cref{sec:complete-benchmarking-results}.
(Although we include results for $\mathcal{M}_1$ from \Cref{fig:valid-expansion-m1} with values $d < 2$ and assuming a unique root, it should be noted that the correct structure of $\mathcal{M}_1$ can only be inferred by selecting $d \geq 2$ and allowing for non-unique roots.)

\begin{wrapfigure}[11]{r}{0.37\textwidth}
    \centering
    \vspace{-30pt}
    \centering
    \begin{tikzpicture}

\begin{axis}[
    at={(0cm, 0cm)}, anchor=south east,
    legend style={nodes={scale=0.5, transform shape}},
    title style={yshift=-1.5ex,align=center},
    title={ActiveMQ (2 clients)\\ \scriptsize{$|\mathcal{M}| = 18$, $|I| = 9$}},
    width=4.3cm,height=3.7cm,
    ymode=log,
    xlabel=\scriptsize{$k$},
    ylabel style={align=center},
    ylabel={\shortstack{\scriptsize{Steps and resets} \\ \scriptsize{per chosen $k$}}},
    xtick={0, 500, ..., 1000},
    ytick={100, 1000, 10000},
    xtick pos=bottom,
    ytick pos=left,
    ymin=100, ymax=1400,
    name=plot1,
]

\addplot[
    color=Green,
    mark=o,
    mark size=1.1pt,
    line width=0.5pt
    ]
    coordinates {
        (25, 1108.44)(50, 959.22)(75, 909.48)(100, 884.61)(125, 869.688)(150, 859.74)(175, 852.6342857142857)(200, 847.305)(225, 843.16)(250, 839.844)(275, 837.1309090909091)(300, 834.87)(325, 832.9569230769231)(350, 831.3171428571428)(375, 829.896)(400, 828.6525)(425, 827.555294117647)(450, 826.58)(475, 825.7073684210526)(500, 824.922)(525, 824.2114285714285)(550, 823.5654545454546)(575, 822.9756521739131)(600, 822.435)(625, 821.9376)(650, 821.4784615384615)(675, 821.0533333333333)(700, 820.6585714285715)(725, 820.2910344827586)(750, 819.948)(775, 819.6270967741935)(800, 819.32625)(825, 819.0436363636363)(850, 818.7776470588235)(875, 818.5268571428571)(900, 818.29)(925, 818.0659459459459)(950, 817.8536842105264)(975, 817.6523076923077)(1000, 817.461)
    };

\addplot[
    color=Orange,
    mark=o,
    mark size=1.1pt,
    line width=0.5pt
    ]
    coordinates {
        (25, 724.64)(50, 418.68)(75, 325.6933333333333)(100, 281.77)(125, 255.816)(150, 238.51333333333332)(175, 226.16571428571427)(200, 217.145)(225, 210.1288888888889)(250, 204.516)(275, 199.92363636363638)(300, 196.09666666666666)(325, 192.85846153846154)(350, 190.08285714285714)(375, 187.72)(400, 185.7375)(425, 183.98823529411766)(450, 182.43333333333334)(475, 181.0421052631579)(500, 179.79)(525, 178.65714285714284)(550, 177.62727272727273)(575, 176.68695652173912)(600, 175.825)(625, 175.032)(650, 174.3)(675, 173.62222222222223)(700, 172.99285714285713)(725, 172.40689655172415)(750, 171.86)(775, 171.3483870967742)(800, 170.86875)(825, 170.4181818181818)(850, 169.99411764705883)(875, 169.59428571428572)(900, 169.21666666666667)(925, 168.85945945945946)(950, 168.52105263157895)(975, 168.2)(1000, 167.895)
    };

\end{axis}

\end{tikzpicture}
    \vspace{-15pt}
    \caption{Impact of $k$ ($d = 0$)}
    \label{fig:experiment-2}
\end{wrapfigure}
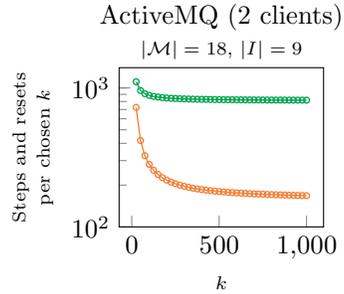\myparagraph{Evaluation.} \textbf{RQ1:} Across all cases, the sampling-based $L^*$ is outperformed by the approach using the expanded \ac{MM} $\mathcal{M}^d$ by a large margin.
\Cref{fig:experiment-2} 
documents the unsurprising finding that for both methods the runtime grows linearly with $k$ (once the effect of the initial $L^*$ step fades out). 

\noindent\textbf{RQ2:} For both methods, the number of actions needed increases mostly exponentially with increasing parameter~$d$. However, the analysis  still is feasible for larger values of $d$ if using expansion $\mathcal{M}^d$ for sample collection. Exceptions to the exponential growth are due to the size of $\mathcal{M}^d$ no longer increasing for larger choices of $d$, rooted in acyclicity (see also the database example in \Cref{sec:practice}).

\noindent \textbf{RQ3:} The size of the input alphabet can induce a massive performance hit for the sampling-based $L^*$ approach, while the size of $\mathcal{M}^d$ remains manageable, as exemplified in the OpenSSH example. The reason is that the former approach needs to maintain a test set for the non-empty words $I^{\leq d}$ which increases significantly in size for a large $|I|$, whereas the expanded machine $\mathcal{M}^d$ eradicates many combinations, e.g., by the stutter-free restriction~(see \Cref{def:no-self-loops}). 

\noindent \textbf{RQ4:} The assumption of a unique root can lead to a vast reduction in the size of $\mathcal{M}^d$, especially in small, cyclic examples, as the transition history of a state becomes irrelevant as soon as the (by assumption unique) root is visited.
In some larger examples, there is also a small size difference w.r.t.~$\mathcal{M}^d$ noticeable, for instance in the ActiveMQ example. However, in most cases, the presence or absence of the assumption has no impact on the number of steps needed.

\section{Towards Practice: When the Order of Tasks Matters}
\label{sec:practice}

For the benefit of full reproducibility and internal validity~\cite{DBLP:conf/icse/SiegmundSA15} and in line with the foundational nature of our contribution, the above empirical studies are fully synthetic. Nevertheless, to obtain a first view on ecological validity, we now complement them by reporting on experiments where real measured delays occur. Specifically, we consider a scenario where there is an intrinsic need for state separations, namely a situation where a system component may offer to perform some tasks but where the order of them being requested determines the overall time to completion, and where it is not clear a~priori which orders are supported at all. So there is a need to learn the supported orders while studying the delays induced. Such situations arise in very many practical settings (e.g., in compiler technology, network management, information retrieval, or verification). For the purpose of a concrete setting, we work with a relational database, holding a set of tables of different sizes.
For technical reasons, it is a~priori unknown whether all such orders are supported. 

\myparagraph{Experimental setup.} Experiments were carried out with PostgreSQL~10.23 on an AMD EPYC 72F3 8-Core CPU with 252~GB of RAM.
The database holds four tables $T_4$ to $T_7$ with sizes ranging from $10^4$ to $10^7$.
The database is orchestrated by a Python wrapper that first collects the order of the joins to perform before finally committing the join order but discards some orders (those starting with $T_4$ then $T_5$ and those ending in $T_5$ then $T_7$, all unknown to the learner upfront).
We always join on the row index, so the joined result has the number of rows of the smaller table involved. To confirm that the logical structure alone has been inferred correctly, we apply the $W$-method~\cite{W-method, W-method-2} with $e=1$. We again use $k = 1000$ and the distribution equality test from~\Cref{sec:evaluation}.

\myparagraph{Results.}
For increasing values of $d$, we report the average delays until the full completion of the transaction (more details of the distributions can be obtained, of course).
\begin{itemize}
\item For $d=0$, the algorithm learns the \ac{MDM} as to be expected, and reports that, regardless of the order of joins, the commit will be delayed with a mean duration of $191$ ms. Note that this choice of $d$ cannot distinguish the effect of different orders, and also does not induce any meaningful averaged mean of the actual delays of the different join orders (the result depending on how often a transition is taken in the true \ac{SUL}).
\item For $d=1$, the last action is decisive for the delay observed until the commit. This delay ranges from a mean $47$ ms observed when $T_7$ is last, to $432$ ms, observed when $T_5$ is last.
\item For $d=2$, the delays observed essentially depend on the last two joins requested, ranging from $21$ ms obtained if the order ends in $T_7$ followed by $T_6$, to $442$ ms obtained if the order ends in $T_4$ followed by $T_5$.
\item For $d=3$, the delays observed stay unchanged from the $d=2$ case because the first binary join behaves the same with respect to the time it takes, regardless of the order of arguments. Still, the size of the final \ac{MDM} $\mathcal{M}^d$ can differ slightly due to stochastic uncertainties in the delay sampling.
\item For $d=4$, the results are as for $d=3$, because the \ac{SUL} is $3$-step confluent. 
\end{itemize}

We report the total runtime, the size of the expanded \ac{MM} $\mathcal{M}^d$, and the size of the resulting minimal \ac{MDM} $\mathcal{D}'$ in \Cref{tab:results-database}.
As one can see, increasing parameter $d$ results in an increased runtime, in line with the findings reported in \Cref{sec:evaluation}, together with a larger number of states, indicating that different delay behaviors caused by different join orders have been discovered successfully. A plateau is reached at $d=3$, as the size of the \ac{MM} $\mathcal{M}^d$ does not change anymore.

\begin{table}[h!]
    \centering
    \caption{Experimental results for database join orders}
    \label{tab:results-database} 
    \setlength{\tabcolsep}{8pt}
    \begin{tabular}{r|rrrrr}\toprule
         $d$ & $0$ & $1$ & $2$ & $3$ & $4$\\
         \midrule
         Runtime in seconds & $\approx 192$ & $\approx 943$ & $\approx 1263$ & $\approx 2469$ & $\approx 2476$ \\
         $|\mathcal{M}^d|$ & $18$ & $33$ & $49$ & $58$ & $58$ \\
         $|\mathcal{D}'|$ & $18$ & $20$ & $29$ & $29$ & $29$ \\ \bottomrule
    \end{tabular}
\end{table}

\section{Conclusion}

This paper has studied the active learning of Mealy machines with stochastic transition delays. To outperform the straightforward adaptation of standard \ac{AAL} algorithms, we have proposed a combination of learning just the \ac{I/O} structure with an efficient way to subsequently collect samples of  transition delays. When facing I/O-equivalent behavior associated with distinct delay characteristics, we have focused on cases with confluent structures of bounded depth, while indicating that the theory behind the approach extends to other settings. The  empirical results reported are very encouraging. 

A notable dimension of future work will be to exploit the cost model underlying the \ac{RPP}, for instance to discriminate between resets and transition traversals based on the true overhead a reset induces, and/or to incorporate the delay information already gathered, thereby achieving a reduction of the overall learning duration incurred in realistic settings.

\paragraph{Acknowledgments.} We thank Tobias Dick from Saarland University for his assistance in conducting the database case study and the fruitful discussions.

\paragraph{Data availability statement.}
The artifact of this paper---a reproduction package for the experiments in \Cref{sec:evaluation} and \Cref{sec:practice}---is available at DOI\:\,\href{https://doi.org/10.6084/m9.figshare.29966641}{\ttfamily\color{blue}10.6084/\allowbreak m9.figshare.29966641}.
The source code is available on \href{https://dgit.cs.uni-saarland.de/gabriel.dengler/al-expect-delays}{\ttfamily\color{blue}https://\allowbreak dgit.\allowbreak cs.\allowbreak uni-\allowbreak saarland.\allowbreak de/\allowbreak gabriel.dengler/\allowbreak al-expect-delays}.

\bibliographystyle{splncs04}
\bibliography{references.bib}

\appendix

\section{Semantics of Mealy Delay Machines}
\label{sec:sema-mdm}

This section spells out the formal semantics of \acp{MDM} by associating a probability measure to  timed output words, whenever a specific input word is fixed. 

An \ac{MM} naturally associates to any input word an output word of the same length. Similarly, an \ac{MDM} $\mathcal{D} = (S, I, O, s_0, \delta, \lambda, D)$  operating on  $w \in I^*$  will induce a probabilitity measure  $\text{Pr}^w$ on the timed output words of $w$'s length, constructed from those of their prefixes. A timed output word of length $n$ is a sequence $\tto{t_1}o_1  \ldots \tto{t_n}o_{n}$ with $t_i\in  \mathbb{R}_0^+$ and $o_i \in O$. We denote the set of all timed output words of length at most $n$ as $\textit{Paths}^{\mathcal{D}\downarrow n}$. (We could equally well include the state identities inside paths, since they are, just like the outputs, uniquely determined given $w$, but this will just clutter the exposition). Obviously, $\textit{Paths}^{\mathcal{D}\downarrow n} \subseteq \textit{Paths}^{\mathcal{D}\downarrow n+1} $.
 
Let $m$ be the length of $w = w_1\ldots w_m$ which we assume fixed. A probability measure $\text{Pr}^w$ on $\textit{Paths}^{\mathcal{D}\downarrow m}$ is defined using a variation of the standard cylinder set construction~(see e.g. \cite{DBLP:conf/birthday/BaierHHK08,DBLP:conf/papm/LopezHK01}) as follows. 
Let $I_1, ..., I_{n}$ be non-empty intervals in $\mathbb{R}_{0}^{+}$ and $o_1, ..., o_n \in O$. 
Then, $C(I_1, o_1, \ldots ,  I_n,o_n)$ denotes the cylinder set consisting of all paths  $\tto{t_1}o_1  \ldots \tto{t_n}o_{n} \in \textit{Paths}^{\mathcal{D}\downarrow n}$ such that $t_i \in I_i$ for all $ i \leq n$.
Let $\mathcal{F}(\textit{Path}^{\mathcal{D}\downarrow m})$ be the smallest $\sigma$-algebra which contains all sets $C(I_1,o_1, \ldots , I_n, o_n)$, where  $I_1, ..., I_n$ ranges over all sequences of non-empty intervals in $\mathbb{R}_0^+$ of length at most $m$, and likewise $o_1, \ldots, o_n$ ranges over all output sequences of length at most~$m$.
The probability measure $\text{Pr}^w$ on $\mathcal{F}(\textit{Path}^{\mathcal{D}\downarrow m})$ is the unique measure defined inductively by $\text{Pr}(C()) = 1$ and for $0 < n < m$,  $
\text{Pr}(C(I_1, o_1, \ldots , I_{n}, o_{n}, I', o')) = 0$ if $\lambda(w_1 \ldots w_{n+1}) \neq o'$ and otherwise it is:
\begin{align*}
    &\textit{Pr}(C( I_1, o_1, \ldots , I_{n},o_{n}))\\
&\cdot (D(\delta(w_1 \ldots w_{n}), w_{n+1})(\sup(I')) - D(\delta(w_1\ldots w_{n}), w_{n+1})(\inf(I')))
\end{align*}
    
The above construction does not account for resets, since resets are not considered inputs. However, we are free to model resets explicitly, by adding a fresh input $\textit{reset}$ to the input set $I$ (and adding reset transitions explicitly to the \ac{MDM} model from each state).

\section{Omitted Proofs}
\label{sec:omitted-proofs}

\begin{proof}[\Cref{lem:io}] We only discuss why I/O equivalence implies bisimilarity, since the converse direction is obvious. 

Consider two MMs $\mathcal{M} = (S, I, O, s_0, \delta, \lambda)$ and $\mathcal{M}' = (S', I, O, s'_0, \delta', \lambda')$, and assume $s_0 \sim_{IO} s'_0$. We define relation $R = \{ (s, s') \mid s \sim_{IO} s'\}$ (for this use of $\sim_{IO}$ to match Definition 1 we work in the disjoint union of the two MMs with arbitrary initial state but define the relation on $S \times S'$). We will show that $R$ fulfills all requirements for being a bisimulation relation between $\mathcal{M}$ and $\mathcal{M}'$ according to Definition 2. First, $s_0\ R\ s'_0$ holds as the initial states are evidently I/O equivalent. Now, consider an arbitrary pair $(s,s') \in R$. For each $i \in I$, the output functions match, i.e., $\lambda(s, i) = \lambda'(s', i)$, as $s\sim_{IO} s'$ (they are functions, so the output letters are each uniquely determined by $i$). It remains to show that the successor states $\delta(s, i)$ and $\delta'(s', i)$ (again, these states are each uniquely determined by $i$) are also pairwise part of relation $R$. Assume this was not the case for some $i$, so we would have that $\delta(s, i)$ and $\delta'(s', i)$ are not I/O equivalent, despite $s\sim_{IO} s'$. So, there is a word $w$ such that $\lambda(\delta(s,i),w) \neq \lambda'(\delta'(s',i), w)$, but, on the other hand, $\lambda(s, iw) = \lambda'(s', iw)$. This is impossible, because the state after $i$ from $s$ is $\delta(s, i)$, and from $s'$ is $\delta'(s', i)$. This concludes the proof.
\end{proof}

\begin{proof}[\Cref{thm:Mstar-inj-bisim}]
    Consider any \ac{MM} $\mathcal{M}$ that fulfills the stutter-free $d$-step confluence criterion with the corresponding minimal (bisimilar) machine $\mathcal{M}'$.
    To show the existence of an injective bisimulation $R$ from $\mathcal
    {M}$ to $\mathcal{M}^d$,
    we remind ourselves that, based on the $d$-step confluence criterion (\Cref{def:d-step-confluence}), a state of the minimal \ac{MM} $s' \in S'$ and a loop-free word $w\in I^{=d}$ starting from $s'$ (in the following we always assume $w$ to be loop-free starting from the referred state) uniquely identifies a state in $\mathcal{M}$. Furthermore, we can also identify a state in $\mathcal{M}$ by an input sequence $w\in I^*$ from the root state $s_0$, where we are interested in the cases $|w| < d$.
    Now, for any state $s\in Q$, we collect the following identifiers:
    \begin{align*}
        \textsf{ids}_s &= \{ \langle s'_p, w\rangle \mid \delta(s_p, w) = s, s_p \sim_{\text{IO}} s'_p \ \ \forall s'_p \in \mathcal{M}', \forall w\in I^{=d} \}\\
        & \qquad \cup \{ \langle s'_0, w\rangle \mid \delta(s_0, w) = s  \ \ \forall w\in I^{<d}\}
    \end{align*}
    Here, we implicitly assume $\sim_{\text{IO}}$ to be applicable for states across different \acp{MM}.
    The first line summarizes the way to identify a state by the corresponding state from $\mathcal{M}'$ and an input sequence of length $d$, while the second line refers to states identified by the root and a word with a length smaller than $d$.
    
    As can be seen, the $\textsf{ids}_s$ are equivalent to the state identifiers of $\mathcal{M}^d$. Thus, we can use the set $\textsf{ids}_s$ to directly relate each state $s\in S$ to the states in $S^d$.
    The resulting injective bisimulation relation is:
    \begin{align*}
        R = \{ (s, s^d_{\textsf{id}_s}) \mid \forall s\in S, \forall \textsf{id}_s \in \textsf{ids}_s \}
    \end{align*}
    $R$ is a valid bisimulation relation: First, the root states match as $s_0\ R\ s^d_{\langle s'_0, \epsilon \rangle}$. Second, the output function matches for all related elements, e.g. $\lambda(s, i) = \lambda^d(s^d_{\textsf{id}_s}, i)$ for all $s\in S, i \in I$, as the output functions of $s$ and $s^d_{\textsf{id}_s}$ are directly inherited from the state $s'$ (where $s'\in S'$ is the corresponding state of $s\in S$ in the minimal \ac{MM} $\mathcal{M}'$). Third, it holds $\delta(s, i)\ R\ \delta(s^d_{\textsf{id}_s}, i)$ (successor in bisimulation) for all $s\in S, i\in I$, as:
    \begin{align*}
    	\left(\delta(s, i), \delta^d(s^d_{s', \langle s'_p, w\rangle}, i)\right) &= \left(\delta(s, i), s^d_{\delta'(s', i), \langle \delta'(s'_p, w_1), w_2...w_{d}i \rangle}\right) \in R \quad \forall w \in I^{=d}\\
    	\left(\delta(s, i), \delta^d(s^d_{s', \langle s'_0, w\rangle}, i)\right) &= \left(\delta(s, i), s^d_{\delta'(s', i), \langle \delta'(s'_0, i), wi \rangle}\right) \in R \qquad \forall w \in I^{<d}
    \end{align*}
    As any state in $\mathcal{M}$ can be uniquely identified by any element of $\textsf{ids}_s$, each identifier set $\textsf{ids}_s$ contains different values, making the relation $R$ injective. This concludes the proof.
\end{proof}

\begin{proof}[\Cref{thm:size-expanded-machine}]
    The construction induces when solely considering all identifiers $\langle s', w\rangle$ with $s'\in S'$ and $w \in I^{=d}$, at maximum, $|S'| \cdot |I|^d$ states for $\mathcal{M}$, one for each source state and input sequence. The tree construction starting from the root state accounts, at maximum, for another $\sum_{j=0}^{d - 1} |I|^j = (|I|^d - 1)/(|I| - 1)$ states. In practice, the actual number of states can be much smaller, e.g., as self-loops do not increase the state space. In total, the number of created states behaves according to $\mathcal{O}(|S'| \cdot |I|^d)$.
\end{proof}

\begin{proof}[\Cref{thm:increasing-d-smaller}]
    We set as the injective bisimulation relation between $\mathcal{M}^d$ and $\mathcal{M}^{d+1}$:
    \begin{align*}
        R &= \{ (s^d_{s', \langle \delta^d(s'_p, j), w\rangle}, s^{d+1}_{s', \langle s'_p, jw \rangle}) \mid s' = \delta'(s_p', jw) \ \ \forall s_p' \in \mathcal{M}', \forall w \in I^{=d}, \forall j \in I \} \\
        & \qquad \cup \{ (s^d_{s', \langle s'_0, w\rangle}, s^{d+1}_{s', \langle s'_0, w \rangle}) \mid s' = \delta'(s_0', w) \ \ \forall w \in I^{\leq d} \}
    \end{align*}
    As before, we are only considering loop-free words. 
    
    We show similarly to the proof of \Cref{thm:Mstar-inj-bisim} that $R$ is a valid bisimulation relation: First, the root states match as $s^d_{\langle s'_0, \epsilon\rangle}\ R\ s^{d+1}_{\langle s'_0, \epsilon \rangle}$. Second, the output function matches for all related elements, as the output function of each state is directly inherited from the state $s'$ of the minimal \ac{MM} $\mathcal{M}'$. Third, also the successors are part of the bisimulation relation after the input $i\in I$:
    \begin{align*}
        &\left(\delta^d(s^d_{s', \langle \delta^d(s'_p, j), w\rangle}, i), \delta^{d+1}(s^{d+1}_{s'\, \langle s'_p, jw \rangle}, i)\right) \\ 
        &\quad = \left(s^d_{\delta'(s', i), \langle \delta^d(s'_p, jw_1), w_2...w_di\rangle}, s^{d+1}_{\delta'(s', i), \langle \delta'(s'_p, j), wi \rangle}\right) \in R \qquad \forall w \in I^{=d} \\
        &\left(\delta^d(s^d_{s', \langle s'_0, w\rangle}, i), \delta^{d+1}(s^{d+1}_{s', \langle s'_0, w \rangle}, i)\right) \\
        &\quad = \left(s^d_{\delta'(s', i), \langle s'_0, wi\rangle}, s^{d+1}_{\delta'(s', i), \langle s'_0, wi\rangle}\right) \in R \qquad \forall w\in I^{< d}\\
        &\left(\delta^d(s^d_{s', \langle s'_0, w\rangle}, i), \delta^{d+1}(s^{d+1}_{s', \langle s'_0, w \rangle}, i)\right) \\
        &\quad = \left(s^d_{\delta'(s', i), \langle \delta'(s'_0, w_1), w_2...w_di\rangle}, s^{d+1}_{\delta'(s', i), \langle s'_0, wi\rangle}\right) \in R \qquad \forall w\in I^{= d}
    \end{align*}
    By construction, $R$ is injective, as the identifier of the state $\mathcal{M}^{d+1}$ uniquely identifies the corresponding state in $\mathcal{M}^d$: The identifier sequence for states in $\mathcal{M}^{d+1}$ is one step longer or equal (when starting from the root) as for the corresponding state in $\mathcal{M}$.
\end{proof}

\begin{proof}[\Cref{thm:feasible-inj-bisim}]
    We provide a construction of such an \ac{MM} $\mathcal{M}^* = (\splitatcommas{S^*, I, O, s^*_0, \delta^*, \lambda^*})$ and show that $\mathcal{M}_j$ for $j\in \{1, ..., n\}$ is injective bisimilar to $\mathcal{M}^*$. We set $S^* = S_1 \times ... \times S_n$ and, for all $(s_1, ..., s_n) \in Q^*$ and $i\in I$:
    \begin{align*}
        \delta((s_1, ..., s_n), i) &= (\delta_1(s_1, i), ..., \delta_n(s_n, i))\\
        \lambda((s_1, ..., s_n), i) &= \lambda_1(s_1, i)
    \end{align*}
    Since all $\mathcal{M}_1, ..., \mathcal{M}_n$ are bisimilar, we only need the output of one of the machines.
    States in $\mathcal{M}^*$ that are not reachable from the initial state $(s_{1,0}, ..., s_{n,0})$ are omitted.
    Thus, it follows directly that the number of resulting states is at maximum $|\mathcal{M}_1| \cdot ...\cdot |\mathcal{M}_n|$.
    
    We can now set up for each $\mathcal{M}_j$ the injective bisimulation relation $R$ between $\mathcal{M}_j$ and $\mathcal{M}^*$:
    \begin{align*}
    	R = \{ \langle s_j, (s_1, ..., s_j, ..., s_n) \rangle \mid \forall s_1 \in S_1, ..., s_n \in S_n \}
    \end{align*}
    $R$ is a valid bisimulation relation: First, the root states match as $s_{j, 0}\ R\ (\splitatcommas{s_{1,0}, ..., s_{n,0}})$. Second, the output functions match as for all words $w\in I^*$, $\lambda_1(w) = ... = \lambda_n(w)$, and thus it holds for the outputs $\lambda_1(s_1, i) = ... = \lambda_n(s_n, i)$ when $s_1, ..., s_n$ have been reached with the same word $w$. Third, it holds $\delta_j(s_j, i)\ R \ \delta^*(\splitatcommas{(s_1, ..., s_n), i})$ for all $i\in I$ as $\delta^*((s_1, ..., s_n), i) = (\delta_1(s_1, i), ..., \delta_n(s_n, i))$.
    The relation $R$ is injective, as for every $(s_1, ..., s_j, ..., s_n) \in S^*$, there is only one $s_j$ matched by the relation $R$.
\end{proof}

\section{Deriving Sample Size for Exponential Distributions}
\label{sec:bounding-relative-error-exponential}

This section is concerned with the derivation of the needed number of samples to derive the sample mean $\overline{X}$ of an exponentially distributed \ac{RV} $X$ with parameter $\lambda$ and mean $E[X] = \mu = 1/\lambda$, such that $\mu$ is estimated with a maximum relative error of $\epsilon$ and a maximum failure probability of $\delta$.

\myparagraph{Exact approach.} It is known that the sum of $k$ independent samples $X_i$ with $1\leq i\leq k$ converges to a Gamma distribution with $\sum_{i=1}^{k} X_i = \Gamma(k, \lambda)$, or equivalently $(1/\mu) \sum_{i=1}^{k} X_i = \Gamma(k, 1)$, and finally:
\begin{align*}
    \frac{1}{\mu k} \sum_{i=1}^{k} X_i = \Gamma(k, k)
\end{align*}
As $\Gamma(k, k)$ has a mean of $1$, we are interested in bounding the failure probability $\delta$ while achieving a relative error bound of $\epsilon$:
\begin{align*}
    \Pr (1 - \epsilon \leq x \leq 1 + \epsilon) \geq 1-\delta \qquad x \sim \Gamma(k, k)
\end{align*}
With $F(x, k)$ being the \ac{CDF} of $\Gamma(k, k)$, the problem can be reformulated to: 
\begin{align*}
    F(1 + \epsilon, k) - F(1 - \epsilon, k) \geq 1-\delta
\end{align*}
As this expression is hard to solve numerically, the easiest approach is to try different values of $k$ until the inequality is fulfilled.

\myparagraph{Approximate approach.} Alternatively, we can make use of the fact that the Gamma distribution (or equivalently the sum of exponential distributions) will converge to a normal distribution for large values of $k$:
\begin{align*}
    \frac{1}{\mu k} \sum_{i=1}^{k} X_i \longrightarrow \mathcal{N}(1, 1/k)
\end{align*}
As before, we are interested in bounding the error:
\begin{align*}
    \Pr (1 - \epsilon \leq x \leq 1 + \epsilon) \geq 1-\delta \qquad x \sim \mathcal{N}(1, 1/k)
\end{align*}
As the normal distribution is, in contrast to the Gamma distribution, symmetrically around its mean, it suffices to determine the locations of the $\delta/2$ and $1 - \delta/2$ quantiles to find the needed $k$. Thus, it follows with $z_{1-\delta/2}$ being the $1 - \delta/2$ quantile of the standard normal distribution:
\begin{align*}
    \frac{1}{\sqrt{k}}\cdot z_{1-\delta/2} \leq \epsilon \qquad \iff \qquad k \geq \left(\frac{z_{1-\delta/2}}{\epsilon}\right)^2
\end{align*}

\myparagraph{Calculation example.} When using $k = 1000$ samples for each edge, then we obtain, for a confidence of $1 - \delta = 0.99$, a maximum relative error of $\epsilon \approx 0.081$. So, with a probability of at least $99$ percent, the estimated mean of an exponential distribution with rate $\mu$ lies in the interval $[(1 - 0.081)\cdot \mu, (1 + 0.081)\cdot \mu]$. Now consider twice $k$ samples from two identical distributions where it is, however, unknown, if they are from the same distribution. We want to guarantee that these distributions are always considered equivalent with confidence $1 - \delta$. In the worst case, we estimate for one distribution as mean the upper limit of the interval, whereas we estimate the lower limit for the other distributions. Thus, as a simple upper bound, two distributions should be considered equal for a relative error up to $(1 + 0.081)/(1 - 0.081) - 1 \approx 0.177$. For simplicity, we round the relative error bound up to $20$ percent in our experiments.

\newpage

\section{Complete Benchmarking Results}
\label{sec:complete-benchmarking-results}

\makeatletter
\setlength{\@fptop}{0pt}
\makeatother

This appendix presents an exhaustive benchmarking of a publicly available test suite for \acp{MM}~\cite{automata-test-suite}.
We stopped an experiment whenever we reached more than $10^8$ executed steps including resets. In this case, a \xmark{} is printed in the corresponding cell. Furthermore, notice that the number of executed actions can differ slightly between assuming a unique root or not despite the size of the expanded \ac{MM} $\mathcal{M}^d$ being the same, as we used a heuristic approach to find a solution to the \ac{RPP} that can produce different results based on the input constellation.

\begin{table}[!htbp]
    \centering
    {\scriptsize
        \begin{tabular}{p{1.2cm}p{1.5cm}|c|x{1.15cm}x{1.15cm}|x{0.7cm}x{1.15cm}x{1cm}|x{0.7cm}x{1.15cm}x{1cm}}
            & & & & & \multicolumn{6}{c}{Using expansion $\mathcal{M}^d$} \\
            \multicolumn{2}{l|}{Experiment} & & \multicolumn{2}{c|}{Adapted $L^*$} & \multicolumn{3}{c|}{\sout{Unique root assumption}} & \multicolumn{3}{c}{Unique root assumption} \\
            $|\mathcal{M}|$ & $|I|$ & $d$ & Inputs & Resets & $|\mathcal{M}^d|$ & Inputs & Resets & $|\mathcal{M}^d|$ & Inputs & Resets \\
            \hline\hline
            \multicolumn{2}{l|}{test\_M1} & $0$ & 12002 & 5996 & 3 & 6012 & 13 & 3 & 6012 & 13 \\
            $|\mathcal{M}| = 5$ & $|I| = 2$ & $1$ & 36000 & 12000 & 7 & 14012 & 2001 & 5 & 12012 & 13 \\
             & & $2$ & 184000 & 40000 & 15 & 32012 & 4001 & 7 & 24012 & 13 \\
             & & $3$ & 448000 & 80000 & 31 & 72012 & 8005 & 7 & 24012 & 13 \\
             & & $4$ & 1056000 & 160000 & 63 & 160020 & 16009 & 7 & 24012 & 13 \\
             & & $5$ & 2432000 & 320000 & 127 & 352036 & 32013 & 7 & 24012 & 13 \\
            \hline
            \multicolumn{2}{l|}{coffeemachine} & $0$ & 56164 & 23998 & 6 & 28229 & 6053 & 6 & 28229 & 6053 \\
            $|\mathcal{M}| = 6$ & $|I| = 4$ & $1$ & 320000 & 96000 & 11 & 54264 & 10053 & 7 & 31229 & 6053 \\
             & & $2$ & 1664000 & 384000 & 21 & 130286 & 22053 & 8 & 47229 & 9053 \\
             & & $3$ & 8192000 & 1536000 & 40 & 235286 & 33053 & 8 & 47229 & 9053 \\
             & & $4$ & 38912000 & 6144000 & 78 & 649286 & 85053 & 8 & 47229 & 9053 \\
             & & $5$ & \xmark & \xmark & 146 & 1211286 & 139053 & 8 & 47229 & 9053 \\
            \hline
            \multicolumn{2}{l|}{naiks} & $0$ & 16014 & 7995 & 4 & 7034 & 22 & 4 & 7034 & 22 \\
            $|\mathcal{M}| = 4$ & $|I| = 2$ & $1$ & 48000 & 16000 & 7 & 14030 & 2001 & 5 & 7042 & 22 \\
             & & $2$ & 128000 & 32000 & 12 & 18034 & 2005 & 6 & 12044 & 22 \\
             & & $3$ & 320000 & 64000 & 18 & 38034 & 4010 & 7 & 12046 & 22 \\
             & & $4$ & 768000 & 128000 & 29 & 53050 & 5017 & 8 & 19046 & 22 \\
             & & $5$ & 1792000 & 256000 & 43 & 94065 & 8021 & 9 & 19046 & 22 \\
            \hline
            \multicolumn{2}{l|}{lee\_yannakakis\_} & $0$ & 12002 & 5996 & 3 & 3024 & 13 & 3 & 3024 & 13 \\
            \multicolumn{2}{l|}{non\_distinguishable} & $1$ & 36000 & 12000 & 4 & 4022 & 1002 & 3 & 3024 & 13 \\
            $|\mathcal{M}| = 3$ & $|I| = 2$ & $2$ & 96000 & 24000 & 5 & 5020 & 1005 & 3 & 3024 & 13 \\
             & & $3$ & 240000 & 48000 & 6 & 6024 & 1009 & 3 & 3024 & 13 \\
             & & $4$ & 576000 & 96000 & 7 & 7032 & 1012 & 3 & 3024 & 13 \\
             & & $5$ & 1344000 & 192000 & 8 & 8036 & 1013 & 3 & 3024 & 13 \\
            \hline
            \multicolumn{2}{l|}{lee\_yannakakis\_} & $0$ & 42093 & 11999 & 6 & 12128 & 47 & 6 & 12128 & 47 \\
            \multicolumn{2}{l|}{distinguishable} & $1$ & 108000 & 24000 & 12 & 24128 & 1002 & 10 & 24128 & 47 \\
            $|\mathcal{M}| = 6$ & $|I| = 2$ & $2$ & 264000 & 48000 & 22 & 48128 & 2004 & 16 & 48128 & 47 \\
             & & $3$ & 624000 & 96000 & 40 & 96128 & 4006 & 24 & 96128 & 47 \\
             & & $4$ & 1440000 & 192000 & 72 & 192128 & 8011 & 32 & 192128 & 47 \\
             & & $5$ & 3264000 & 384000 & 128 & 272116 & 16017 & 32 & 192128 & 47 \\
            \hline
            \multicolumn{2}{l|}{cacm} & $0$ & 12009 & 5996 & 3 & 4025 & 17 & 3 & 4025 & 17 \\
            $|\mathcal{M}| = 3$ & $|I| = 2$ & $1$ & 36000 & 12000 & 5 & 8033 & 1002 & 4 & 8033 & 17 \\
             & & $2$ & 96000 & 24000 & 8 & 12021 & 2003 & 5 & 8035 & 17 \\
             & & $3$ & 240000 & 48000 & 12 & 20035 & 2009 & 6 & 14037 & 17 \\
             & & $4$ & 576000 & 96000 & 18 & 32037 & 4013 & 7 & 14037 & 17 \\
             & & $5$ & 1344000 & 192000 & 26 & 48049 & 4016 & 8 & 22037 & 17 \\
            \hline
        \end{tabular}
    }

    \caption{Benchmarking results for small test examples}
    \label{tab:benchmarking-test-examples}
\end{table}

\newpage

\begin{table}[!htbp]
    \centering
    {\scriptsize
        \begin{tabular}{p{1.2cm}p{1.5cm}|c|x{1.15cm}x{1.15cm}|x{0.7cm}x{1.15cm}x{1cm}|x{0.7cm}x{1.15cm}x{1cm}}
            & & & & & \multicolumn{6}{c}{Using expansion $\mathcal{M}^d$} \\
            \multicolumn{2}{l|}{Experiment} & & \multicolumn{2}{c|}{Adapted $L^*$} & \multicolumn{3}{c|}{\sout{Unique root assumption}} & \multicolumn{3}{c}{Unique root assumption} \\
            $|\mathcal{M}|$ & $|I|$ & $d$ & Inputs & Resets & $|\mathcal{M}^d|$ & Inputs & Resets & $|\mathcal{M}^d|$ & Inputs & Resets \\
            \hline\hline
            \multicolumn{2}{l|}{QUICprotocol} & $0$ & 100343 & 35013 & 7 & 42307 & 7126 & 7 & 42307 & 7126 \\
            \multicolumn{2}{l|}{with0RTT} & $1$ & 675000 & 175000 & 24 & 139455 & 10173 & 24 & 139455 & 10173 \\
            $|\mathcal{M}| = 7$ & $|I| = 5$ & $2$ & 4250000 & 875000 & 71 & 469769 & 28070 & 71 & 469769 & 28070 \\
             & & $3$ & 25625000 & 4375000 & 209 & 1330886 & 70134 & 209 & 1327800 & 73220 \\
             & & $4$ & \xmark & \xmark & 590 & 4455946 & 218196 & 590 & 4431932 & 242210 \\
             & & $5$ & \xmark & \xmark & 1710 & 12910986 & 582227 & 1710 & 12896980 & 596233 \\
            \hline
            \multicolumn{2}{l|}{QUICprotocol} & $0$ & 40044 & 19984 & 5 & 18110 & 4006 & 5 & 18110 & 4006 \\
            \multicolumn{2}{l|}{without0RTT} & $1$ & 240000 & 80000 & 15 & 42120 & 9022 & 15 & 42120 & 9022 \\
            $|\mathcal{M}| = 5$ & $|I| = 4$ & $2$ & 1280000 & 320000 & 32 & 91184 & 15054 & 32 & 91184 & 15054 \\
             & & $3$ & 6400000 & 1280000 & 65 & 227216 & 32074 & 65 & 227216 & 32074 \\
             & & $4$ & 30720000 & 5120000 & 135 & 516240 & 65081 & 135 & 516240 & 65081 \\
             & & $5$ & \xmark & \xmark & 272 & 1146240 & 127081 & 272 & 1146240 & 127081 \\
            \hline
        \end{tabular}
    }
    \caption{Benchmarking results for QUIC models}
    \label{tab:raw_results_QUIC}
\end{table}

\begin{table}[htbp!]
    \centering
        {\scriptsize
        \begin{tabular}{p{1.2cm}p{1.5cm}|c|x{1.15cm}x{1.15cm}|x{0.7cm}x{1.15cm}x{1cm}|x{0.7cm}x{1.15cm}x{1cm}}
            & & & & & \multicolumn{6}{c}{Using expansion $\mathcal{M}^d$} \\
            \multicolumn{2}{l|}{Experiment} & & \multicolumn{2}{c|}{Adapted $L^*$} & \multicolumn{3}{c|}{\sout{Unique root assumption}} & \multicolumn{3}{c}{Unique root assumption} \\
            $|\mathcal{M}|$ & $|I|$ & $d$ & Inputs & Resets & $|\mathcal{M}^d|$ & Inputs & Resets & $|\mathcal{M}^d|$ & Inputs & Resets \\
            \hline\hline
            \multicolumn{2}{l|}{TCP\_Windows8\_} & $0$ & 451474 & 129988 & 13 & 191112 & 41577 & 13 & 191112 & 41577 \\
            \multicolumn{2}{l|}{Client} & $1$ & 5800000 & 1300000 & 26 & 371329 & 76577 & 22 & 367185 & 72577 \\
            $|\mathcal{M}| = 13$ & $|I| = 10$ & $2$ & 71000000 & 13000000 & 55 & 1029040 & 187492 & 43 & 964880 & 175492 \\
             & & $3$ & \xmark & \xmark & 112 & 2310380 & 370577 & 76 & 1914220 & 322577 \\
            \hline
            \multicolumn{2}{l|}{TCP\_Windows8\_} & $0$ & 3000500 & 496932 & 38 & 826054 & 79930 & 38 & 826054 & 79930 \\
            \multicolumn{2}{l|}{Server} & $1$ & 43264000 & 6422000 & 175 & 2727659 & 245810 & 175 & 2727659 & 245810 \\
            $|\mathcal{M}| = 38$ & $|I| = 13$ & $2$ & \xmark & \xmark & 540 & 9358620 & 712750 & 540 & 9358620 & 712750 \\
             & & $3$ & \xmark & \xmark & 1876 & 38731025 & 2669822 & 1876 & 38730029 & 2670818 \\
            \hline
            \multicolumn{2}{l|}{TCP\_Linux\_Server} & $0$ & 4890221 & 689529 & 57 & 984345 & 76509 & 57 & 984345 & 76509 \\
            $|\mathcal{M}| = 57$ & $|I| = 12$ & $1$ & 64080000 & 8208000 & 250 & 4121678 & 329843 & 250 & 4121678 & 329843 \\
             & & $2$ & \xmark & \xmark & 816 & 14504719 & 989778 & 816 & 14504719 & 989778 \\
             & & $3$ & \xmark & \xmark & 3009 & 59044755 & 3707765 & 3009 & 59044755 & 3707765 \\
            \hline
            \multicolumn{2}{l|}{TCP\_FreeBSD\_Client} & $0$ & 411309 & 119979 & 12 & 132927 & 33492 & 12 & 132927 & 33492 \\
            $|\mathcal{M}| = 12$ & $|I| = 10$ & $1$ & 5300000 & 1200000 & 20 & 292672 & 67619 & 20 & 292672 & 67619 \\
             & & $2$ & 65000000 & 12000000 & 29 & 501792 & 109397 & 29 & 501792 & 109397 \\
             & & $3$ & \xmark & \xmark & 40 & 793047 & 157492 & 40 & 793047 & 157492 \\
            \hline
            \multicolumn{2}{l|}{TCP\_FreeBSD\_Server} & $0$ & 5264093 & 721093 & 55 & 1084725 & 78344 & 55 & 1084725 & 78344 \\
            $|\mathcal{M}| = 55$ & $|I| = 13$ & $1$ & 75712000 & 9295000 & 277 & 5191973 & 349095 & 277 & 5191973 & 349095 \\
             & & $2$ & \xmark & \xmark & 953 & 18597638 & 1078072 & 953 & 18597638 & 1078072 \\
             & & $3$ & \xmark & \xmark & 3764 & 83240186 & 4316627 & 3764 & 83240186 & 4316627 \\
            \hline
            \multicolumn{2}{l|}{TCP\_Linux\_Client} & $0$ & 532543 & 150102 & 15 & 163894 & 36864 & 15 & 163894 & 36864 \\
            $|\mathcal{M}| = 15$ & $|I| = 10$ & $1$ & 6800000 & 1500000 & 29 & 419753 & 83768 & 29 & 419753 & 83768 \\
             & & $2$ & 83000000 & 15000000 & 54 & 816767 & 150755 & 54 & 816767 & 150755 \\
             & & $3$ & \xmark & \xmark & 103 & 1965138 & 317848 & 103 & 1965138 & 317848 \\
            \hline
        \end{tabular}
    }
    \caption{Benchmarking results for TCP examples}
    \label{tab:raw_results_tcp}
\end{table}

\newpage

\begin{table}[!htbp]
    \centering
        {\scriptsize
        \begin{tabular}{p{1.2cm}p{1.5cm}|c|x{1.15cm}x{1.15cm}|x{0.7cm}x{1.15cm}x{1cm}|x{0.7cm}x{1.15cm}x{1cm}}
            & & & & & \multicolumn{6}{c}{Using expansion $\mathcal{M}^d$} \\
            \multicolumn{2}{l|}{Experiment} & & \multicolumn{2}{c|}{Adapted $L^*$} & \multicolumn{3}{c|}{\sout{Unique root assumption}} & \multicolumn{3}{c}{Unique root assumption} \\
            $|\mathcal{M}|$ & $|I|$ & $d$ & Inputs & Resets & $|\mathcal{M}^d|$ & Inputs & Resets & $|\mathcal{M}^d|$ & Inputs & Resets \\
            \hline\hline
            \multicolumn{2}{l|}{DropBear} & $0$ & 1056102 & 220971 & 17 & 519773 & 92784 & 17 & 519773 & 92784 \\
            $|\mathcal{M}| = 17$ & $|I| = 13$ & $1$ & 16562000 & 2873000 & 129 & 1638389 & 236696 & 129 & 1638389 & 236696 \\
             & & $2$ & \xmark & \xmark & 398 & 4046296 & 483026 & 398 & 4046296 & 483026 \\
             & & $3$ & \xmark & \xmark & 941 & 8632502 & 879777 & 941 & 8632502 & 879777 \\
            \hline
            \multicolumn{2}{l|}{BitVise} & $0$ & 6146789 & 865015 & 66 & 2302927 & 297984 & 66 & 2302927 & 297984 \\
            $|\mathcal{M}| = 66$ & $|I| = 13$ & $1$ & \xmark & \xmark & 460 & 6936344 & 765624 & 460 & 6936344 & 765624 \\
             & & $2$ & \xmark & \xmark & 1497 & 20349851 & 1995293 & 1497 & 20349851 & 1995293 \\
             & & $3$ & \xmark & \xmark & 4250 & 59621666 & 5294902 & 4250 & 59621666 & 5294902 \\
            \hline
            \multicolumn{2}{l|}{OpenSSH} & $0$ & 3226857 & 684790 & 31 & 1397387 & 254145 & 31 & 1397387 & 254145 \\
            $|\mathcal{M}| = 31$ & $|I| = 22$ & $1$ & \xmark & \xmark & 313 & 2917680 & 500840 & 313 & 2917680 & 500840 \\
             & & $2$ & \xmark & \xmark & 870 & 6972487 & 1031387 & 870 & 6972487 & 1031387 \\
             & & $3$ & \xmark & \xmark & 2021 & 15582811 & 1999494 & 2021 & 15582811 & 1999494 \\
            \hline
        \end{tabular}
    }

    \caption{Benchmarking results for SSH models}
    \label{tab:raw_results_ssh}
\end{table}

\begin{table}[!htbp]
    \centering
        {\scriptsize
        \begin{tabular}{p{1.2cm}p{1.5cm}|c|x{1.15cm}x{1.15cm}|x{0.7cm}x{1.15cm}x{1cm}|x{0.7cm}x{1.15cm}x{1cm}}
            & & & & & \multicolumn{6}{c}{Using expansion $\mathcal{M}^d$} \\
            \multicolumn{2}{l|}{Experiment} & & \multicolumn{2}{c|}{Adapted $L^*$} & \multicolumn{3}{c|}{\sout{Unique root assumption}} & \multicolumn{3}{c}{Unique root assumption} \\
            $|\mathcal{M}|$ & $|I|$ & $d$ & Inputs & Resets & $|\mathcal{M}^d|$ & Inputs & Resets & $|\mathcal{M}^d|$ & Inputs & Resets \\
            \hline\hline
            \multicolumn{2}{l|}{learnresult\_old\_device-} & $0$ & 50055 & 19985 & 4 & 13229 & 101 & 4 & 13229 & 101 \\
            \multicolumn{2}{l|}{simple\_fix} & $1$ & 350000 & 100000 & 10 & 33303 & 1017 & 8 & 33271 & 101 \\
            $|\mathcal{M}| = 4$ & $|I| = 5$ & $2$ & 2250000 & 500000 & 24 & 81326 & 3027 & 17 & 73285 & 101 \\
             & & $3$ & 13750000 & 2500000 & 59 & 235304 & 7048 & 40 & 220285 & 101 \\
            \hline
            \multicolumn{2}{l|}{learnresult\_old\_} & $0$ & 614899 & 176517 & 22 & 228281 & 34034 & 22 & 228281 & 34034 \\
            \multicolumn{2}{l|}{500\_10-15\_fix} & $1$ & 6144000 & 1408000 & 120 & 996778 & 153039 & 120 & 996778 & 153039 \\
            $|\mathcal{M}| = 22$ & $|I| = 8$ & $2$ & 60416000 & 11264000 & 464 & 3754690 & 507235 & 464 & 3754690 & 507235 \\
             & & $3$ & \xmark & \xmark & 1540 & 13662618 & 1597404 & 1540 & 13662618 & 1597404 \\
            \hline
            \multicolumn{2}{l|}{learnresult\_new\_} & $0$ & 30050 & 14990 & 3 & 6165 & 76 & 3 & 6165 & 76 \\
            \multicolumn{2}{l|}{device-simple\_fix} & $1$ & 225000 & 75000 & 7 & 18193 & 1017 & 6 & 18193 & 76 \\
            $|\mathcal{M}| = 3$ & $|I| = 5$ & $2$ & 1500000 & 375000 & 16 & 36201 & 3027 & 12 & 34201 & 76 \\
             & & $3$ & 9375000 & 1875000 & 35 & 104215 & 5055 & 24 & 98201 & 76 \\
            \hline
            \multicolumn{2}{l|}{learnresult\_new\_x} & $0$ & 192788 & 64026 & 8 & 63014 & 6141 & 8 & 63014 & 6141 \\
            \multicolumn{2}{l|}{W-method\_fix} & $1$ & 1984000 & 512000 & 35 & 208416 & 18159 & 33 & 198421 & 14252 \\
            $|\mathcal{M}| = 8$ & $|I| = 8$ & $2$ & 19968000 & 4096000 & 107 & 651727 & 34317 & 97 & 620624 & 34317 \\
             & & $3$ & \xmark & \xmark & 331 & 2323554 & 140343 & 301 & 2182560 & 128382 \\
            \hline
            \multicolumn{2}{l|}{learnresult\_new\_Rand\_} & $0$ & 313673 & 88113 & 11 & 133652 & 8228 & 11 & 133652 & 8228 \\
            \multicolumn{2}{l|}{500\_10-15\_MC\_fix} & $1$ & 3136000 & 704000 & 57 & 459814 & 30262 & 53 & 445659 & 22375 \\
            $|\mathcal{M}| = 11$ & $|I| = 8$ & $2$ & 30720000 & 5632000 & 189 & 1511440 & 80389 & 166 & 1395220 & 58448 \\
             & & $3$ & \xmark & \xmark & 593 & 5551475 & 230460 & 520 & 5203272 & 198538 \\
            \hline
        \end{tabular}
    }

    \caption{Benchmarking results for e.dentifier2}
    \label{tab:raw_results_eidentifier}
\end{table}

\newpage

\begin{table}[!htbp]
    \centering
    {\scriptsize
        \begin{tabular}{p{1.2cm}p{1.5cm}|c|x{1.15cm}x{1.15cm}|x{0.7cm}x{1.15cm}x{1cm}|x{0.7cm}x{1.15cm}x{1cm}}
            & & & & & \multicolumn{6}{c}{Using expansion $\mathcal{M}^d$} \\
            \multicolumn{2}{l|}{Experiment} & & \multicolumn{2}{c|}{Adapted $L^*$} & \multicolumn{3}{c|}{\sout{Unique root assumption}} & \multicolumn{3}{c}{Unique root assumption} \\
            $|\mathcal{M}|$ & $|I|$ & $d$ & Inputs & Resets & $|\mathcal{M}^d|$ & Inputs & Resets & $|\mathcal{M}^d|$ & Inputs & Resets \\
            \hline\hline
            \multicolumn{2}{l|}{Volksbank\_learnresult\_} & $0$ & 337050 & 97916 & 7 & 81436 & 1374 & 7 & 81436 & 1374 \\
            \multicolumn{2}{l|}{MAESTRO\_fix} & $1$ & 6076000 & 1372000 & 36 & 205001 & 1374 & 36 & 205001 & 1374 \\
            $|\mathcal{M}| = 7$ & $|I| = 14$ & $2$ & \xmark & \xmark & 119 & 1265611 & 10399 & 119 & 1265611 & 10399 \\
             & & $3$ & \xmark & \xmark & 499 & 5084503 & 34701 & 499 & 5084503 & 34701 \\
            \hline
            \multicolumn{2}{l|}{Rabo\_learnresult\_} & $0$ & 286080 & 89925 & 6 & 150865 & 1352 & 6 & 150865 & 1352 \\
            \multicolumn{2}{l|}{SecureCode\_Aut\_fix} & $1$ & 5625000 & 1350000 & 58 & 506518 & 1421 & 58 & 506518 & 1421 \\
            $|\mathcal{M}| = 6$ & $|I| = 15$ & $2$ & \xmark & \xmark & 194 & 2576834 & 13442 & 194 & 2576834 & 13442 \\
             & & $3$ & \xmark & \xmark & 882 & 17605065 & 66619 & 882 & 17605065 & 66619 \\
            \hline
            \multicolumn{2}{l|}{4\_learnresult\_} & $0$ & 140658 & 55958 & 4 & 62109 & 1170 & 4 & 62109 & 1170 \\
            \multicolumn{2}{l|}{SecureCode Aut\_fix} & $1$ & 2744000 & 784000 & 29 & 146592 & 1365 & 29 & 146592 & 1365 \\
            $|\mathcal{M}| = 4$ & $|I| = 14$ & $2$ & 49392000 & 10976000 & 80 & 950038 & 12385 & 80 & 950038 & 12385 \\
             & & $3$ & \xmark & \xmark & 384 & 4012765 & 51538 & 384 & 4012765 & 51538 \\
            \hline
            \multicolumn{2}{l|}{ASN\_learnresult\_} & $0$ & 252938 & 83930 & 6 & 128483 & 1178 & 6 & 128483 & 1178 \\
            \multicolumn{2}{l|}{MAESTRO\_fix} & $1$ & 4704000 & 1176000 & 55 & 428375 & 1365 & 55 & 428375 & 1365 \\
            $|\mathcal{M}| = 6$ & $|I| = 14$ & $2$ & 82320000 & 16464000 & 182 & 2189130 & 12385 & 182 & 2189130 & 12385 \\
             & & $3$ & \xmark & \xmark & 860 & 12183466 & 62531 & 860 & 12183466 & 62531 \\
            \hline
            \multicolumn{2}{l|}{learnresult\_fix} & $0$ & 466575 & 134880 & 9 & 70622 & 2027 & 9 & 70622 & 2027 \\
            $|\mathcal{M}| = 9$ & $|I| = 15$ & $1$ & 9000000 & 2025000 & 35 & 250420 & 2772 & 35 & 250420 & 2772 \\
             & & $2$ & \xmark & \xmark & 110 & 614839 & 10659 & 110 & 614839 & 10659 \\
             & & $3$ & \xmark & \xmark & 300 & 1835242 & 47169 & 300 & 1835242 & 47169 \\
            \hline
            \multicolumn{2}{l|}{10\_learnresult\_} & $0$ & 252938 & 83930 & 6 & 128483 & 1178 & 6 & 128483 & 1178 \\
            \multicolumn{2}{l|}{MasterCard\_fix} & $1$ & 4704000 & 1176000 & 55 & 428375 & 1365 & 55 & 428375 & 1365 \\
            $|\mathcal{M}| = 6$ & $|I| = 14$ & $2$ & 82320000 & 16464000 & 182 & 2189130 & 12385 & 182 & 2189130 & 12385 \\
             & & $3$ & \xmark & \xmark & 860 & 12183466 & 62531 & 860 & 12183466 & 62531 \\
            \hline
            \multicolumn{2}{l|}{ASN\_learnresult\_} & $0$ & 140658 & 55958 & 4 & 62109 & 1170 & 4 & 62109 & 1170 \\
            \multicolumn{2}{l|}{SecureCode Aut\_fix} & $1$ & 2744000 & 784000 & 29 & 146592 & 1365 & 29 & 146592 & 1365 \\
            $|\mathcal{M}| = 4$ & $|I| = 14$ & $2$ & 49392000 & 10976000 & 80 & 950038 & 12385 & 80 & 950038 & 12385 \\
             & & $3$ & \xmark & \xmark & 384 & 4012765 & 51538 & 384 & 4012765 & 51538 \\
            \hline
            \multicolumn{2}{l|}{4\_learnresult\_PIN\_fix} & $0$ & 252938 & 83930 & 6 & 128483 & 1178 & 6 & 128483 & 1178 \\
            $|\mathcal{M}| = 6$ & $|I| = 14$ & $1$ & 4704000 & 1176000 & 55 & 428375 & 1365 & 55 & 428375 & 1365 \\
             & & $2$ & 82320000 & 16464000 & 182 & 2189130 & 12385 & 182 & 2189130 & 12385 \\
             & & $3$ & \xmark & \xmark & 860 & 12183466 & 62531 & 860 & 12183466 & 62531 \\
            \hline
            \multicolumn{2}{l|}{1\_learnresult\_} & $0$ & 210930 & 74940 & 5 & 106497 & 1197 & 5 & 106497 & 1197 \\
            \multicolumn{2}{l|}{MasterCard\_fix} & $1$ & 4275000 & 1125000 & 45 & 264049 & 1421 & 45 & 264049 & 1421 \\
            $|\mathcal{M}| = 5$ & $|I| = 15$ & $2$ & 81000000 & 16875000 & 127 & 1767225 & 12448 & 127 & 1767225 & 12448 \\
             & & $3$ & \xmark & \xmark & 644 & 8079090 & 54622 & 644 & 8079090 & 54622 \\
            \hline
            \multicolumn{2}{l|}{4\_learnresult\_} & $0$ & 252938 & 83930 & 6 & 128483 & 1178 & 6 & 128483 & 1178 \\
            \multicolumn{2}{l|}{MAESTRO\_fix} & $1$ & 4704000 & 1176000 & 55 & 428375 & 1365 & 55 & 428375 & 1365 \\
            $|\mathcal{M}| = 6$ & $|I| = 14$ & $2$ & 82320000 & 16464000 & 182 & 2189130 & 12385 & 182 & 2189130 & 12385 \\
             & & $3$ & \xmark & \xmark & 860 & 12183466 & 62531 & 860 & 12183466 & 62531 \\
            \hline
            \multicolumn{2}{l|}{Rabo\_learnresult\_} & $0$ & 252938 & 83930 & 6 & 128483 & 1178 & 6 & 128483 & 1178 \\
            \multicolumn{2}{l|}{MAESTRO\_fix} & $1$ & 4704000 & 1176000 & 55 & 428375 & 1365 & 55 & 428375 & 1365 \\
            $|\mathcal{M}| = 6$ & $|I| = 14$ & $2$ & 82320000 & 16464000 & 182 & 2189130 & 12385 & 182 & 2189130 & 12385 \\
             & & $3$ & \xmark & \xmark & 860 & 12183466 & 62531 & 860 & 12183466 & 62531 \\
            \hline
        \end{tabular}
    }
    \caption{Benchmarking results for collection of bank cards}
    \label{tab:benchmarking-bankcard}
\end{table}

\newpage

\begin{table}[!htbp]
    \centering
    {\scriptsize
        \begin{tabular}{p{1.2cm}p{1.5cm}|c|x{1.15cm}x{1.15cm}|x{0.7cm}x{1.15cm}x{1cm}|x{0.7cm}x{1.15cm}x{1cm}}
            & & & & & \multicolumn{6}{c}{Using expansion $\mathcal{M}^d$} \\
            \multicolumn{2}{l|}{Experiment} & & \multicolumn{2}{c|}{Adapted $L^*$} & \multicolumn{3}{c|}{\sout{Unique root assumption}} & \multicolumn{3}{c}{Unique root assumption} \\
            $|\mathcal{M}|$ & $|I|$ & $d$ & Inputs & Resets & $|\mathcal{M}^d|$ & Inputs & Resets & $|\mathcal{M}^d|$ & Inputs & Resets \\
            \hline\hline
            \multicolumn{2}{l|}{ActiveMQ\_\_invalid} & $0$ & 143473 & 54956 & 5 & 69012 & 606 & 5 & 69012 & 606 \\
            $|\mathcal{M}| = 5$ & $|I| = 11$ & $1$ & 2178000 & 605000 & 30 & 187092 & 1101 & 10 & 187092 & 606 \\
             & & $2$ & 30613000 & 6655000 & 85 & 793958 & 7131 & 20 & 473108 & 606 \\
             & & $3$ & \xmark & \xmark & 320 & 2705688 & 19256 & 40 & 1145108 & 606 \\
            \hline
            \multicolumn{2}{l|}{mosquitto\_\_invalid} & $0$ & 66308 & 32978 & 3 & 32565 & 364 & 3 & 32565 & 364 \\
            $|\mathcal{M}| = 3$ & $|I| = 11$ & $1$ & 1089000 & 363000 & 16 & 56853 & 1101 & 4 & 56583 & 364 \\
             & & $2$ & 15972000 & 3993000 & 35 & 320897 & 7131 & 5 & 86585 & 364 \\
             & & $3$ & \xmark & \xmark & 132 & 674789 & 13246 & 6 & 122585 & 364 \\
            \hline
            \multicolumn{2}{l|}{mosquitto\_\_single\_client} & $0$ & 352869 & 109901 & 10 & 110782 & 24631 & 10 & 110782 & 24631 \\
            $|\mathcal{M}| = 10$ & $|I| = 11$ & $1$ & 5082000 & 1210000 & 27 & 405094 & 75631 & 27 & 405094 & 75631 \\
             & & $2$ & 69212000 & 13310000 & 78 & 1440166 & 228631 & 78 & 1440166 & 228631 \\
             & & $3$ & \xmark & \xmark & 231 & 5004166 & 687631 & 231 & 5004166 & 687631 \\
            \hline
            \multicolumn{2}{l|}{VerneMQ\_\_two\_client\_} & $0$ & 598962 & 153410 & 17 & 144198 & 2194 & 17 & 144198 & 2194 \\
            \multicolumn{2}{l|}{will\_retain} & $1$ & 6237000 & 1377000 & 95 & 756356 & 3875 & 88 & 756356 & 2194 \\
            $|\mathcal{M}| = 17$ & $|I| = 9$ & $2$ & 68526000 & 12393000 & 467 & 3859491 & 14441 & 411 & 3756464 & 2194 \\
             & & $3$ & \xmark & \xmark & 2346 & 20548404 & 55411 & 2001 & 19834843 & 15191 \\
            \hline
            \multicolumn{2}{l|}{VerneMQ\_\_two\_client} & $0$ & 504801 & 143865 & 16 & 85296 & 4257 & 16 & 85296 & 4257 \\
            $|\mathcal{M}| = 16$ & $|I| = 9$ & $1$ & 5832000 & 1296000 & 61 & 336310 & 18292 & 58 & 332278 & 17344 \\
             & & $2$ & 64152000 & 11664000 & 227 & 1352235 & 70385 & 215 & 1315305 & 67497 \\
             & & $3$ & \xmark & \xmark & 850 & 5404624 & 264603 & 806 & 5217721 & 258693 \\
            \hline
            \multicolumn{2}{l|}{hbmqtt\_\_non\_clean} & $0$ & 222779 & 60039 & 10 & 29270 & 4118 & 10 & 29270 & 4118 \\
            $|\mathcal{M}| = 10$ & $|I| = 6$ & $1$ & 1692000 & 360000 & 21 & 67578 & 8200 & 20 & 66605 & 7245 \\
             & & $2$ & 12312000 & 2160000 & 44 & 150945 & 15291 & 40 & 145946 & 14316 \\
             & & $3$ & 86832000 & 12960000 & 90 & 316975 & 29351 & 79 & 288988 & 25366 \\
            \hline
            \multicolumn{2}{l|}{hbmqtt\_\_invalid} & $0$ & 66308 & 32978 & 3 & 4763 & 1222 & 3 & 4763 & 1222 \\
            $|\mathcal{M}| = 3$ & $|I| = 11$ & $1$ & 1089000 & 363000 & 4 & 4803 & 1222 & 3 & 4763 & 1222 \\
             & & $2$ & 15972000 & 3993000 & 5 & 8805 & 2222 & 3 & 4763 & 1222 \\
             & & $3$ & \xmark & \xmark & 6 & 8805 & 2222 & 3 & 4763 & 1222 \\
            \hline
            \multicolumn{2}{l|}{ActiveMQ\_\_two\_client\_} & $0$ & 654848 & 162613 & 18 & 165292 & 2603 & 18 & 165292 & 2603 \\
            \multicolumn{2}{l|}{will\_retain} & $1$ & 6804000 & 1458000 & 103 & 900522 & 4242 & 96 & 900522 & 2603 \\
            $|\mathcal{M}| = 18$ & $|I| = 9$ & $2$ & 74358000 & 13122000 & 513 & 4613907 & 14621 & 457 & 4522952 & 2603 \\
             & & $3$ & \xmark & \xmark & 2589 & 24228939 & 56464 & 2235 & 23431066 & 7593 \\
            \hline
            \multicolumn{2}{l|}{VerneMQ\_\_simple} & $0$ & 42112 & 20986 & 3 & 17248 & 148 & 3 & 17248 & 148 \\
            $|\mathcal{M}| = 3$ & $|I| = 7$ & $1$ & 441000 & 147000 & 10 & 29352 & 1037 & 4 & 29262 & 148 \\
             & & $2$ & 4116000 & 1029000 & 20 & 98375 & 4055 & 5 & 44264 & 148 \\
             & & $3$ & 36015000 & 7203000 & 51 & 194336 & 7101 & 6 & 62264 & 148 \\
            \hline
            \multicolumn{2}{l|}{emqtt\_\_two\_client\_} & $0$ & 654848 & 162613 & 18 & 165292 & 2603 & 18 & 165292 & 2603 \\
            \multicolumn{2}{l|}{will\_retain} & $1$ & 6804000 & 1458000 & 103 & 900522 & 4242 & 96 & 900522 & 2603 \\
            $|\mathcal{M}| = 18$ & $|I| = 9$ & $2$ & 74358000 & 13122000 & 513 & 4613907 & 14621 & 457 & 4520088 & 5467 \\
             & & $3$ & \xmark & \xmark & 2589 & 24229672 & 55731 & 2235 & 23415082 & 23577 \\
            \hline
            \multicolumn{2}{l|}{emqtt\_\_non\_clean} & $0$ & 312298 & 71959 & 12 & 48004 & 494 & 12 & 48004 & 494 \\
            $|\mathcal{M}| = 12$ & $|I| = 6$ & $1$ & 2304000 & 432000 & 35 & 174314 & 2017 & 32 & 174314 & 494 \\
             & & $2$ & 16416000 & 2592000 & 108 & 608484 & 6051 & 93 & 590484 & 494 \\
             & & $3$ & \xmark & \xmark & 336 & 2040538 & 18105 & 275 & 1926526 & 494 \\
            \hline
            \multicolumn{2}{l|}{ActiveMQ\_\_simple} & $0$ & 56147 & 27979 & 4 & 22300 & 6002 & 4 & 22300 & 6002 \\
            $|\mathcal{M}| = 4$ & $|I| = 7$ & $1$ & 588000 & 196000 & 16 & 43400 & 9086 & 16 & 43400 & 9086 \\
             & & $2$ & 5488000 & 1372000 & 34 & 148572 & 27146 & 34 & 148572 & 27146 \\
             & & $3$ & 48020000 & 9604000 & 88 & 358575 & 54190 & 88 & 358575 & 54190 \\
            \hline
            \multicolumn{2}{l|}{VerneMQ\_\_invalid} & $0$ & 66308 & 32978 & 3 & 32565 & 364 & 3 & 32565 & 364 \\
            $|\mathcal{M}| = 3$ & $|I| = 11$ & $1$ & 1089000 & 363000 & 16 & 56853 & 1101 & 4 & 56583 & 364 \\
             & & $2$ & 15972000 & 3993000 & 35 & 320897 & 7131 & 5 & 86585 & 364 \\
             & & $3$ & \xmark & \xmark & 132 & 674789 & 13246 & 6 & 122585 & 364 \\
            \hline
            \multicolumn{2}{l|}{hbmqtt\_\_single\_client} & $0$ & 352869 & 109901 & 10 & 83236 & 16768 & 10 & 83236 & 16768 \\
            $|\mathcal{M}| = 10$ & $|I| = 11$ & $1$ & 5082000 & 1210000 & 27 & 295596 & 50768 & 27 & 295596 & 50768 \\
             & & $2$ & 69212000 & 13310000 & 78 & 1033668 & 152768 & 78 & 1033668 & 152768 \\
             & & $3$ & \xmark & \xmark & 231 & 3553668 & 458768 & 231 & 3553668 & 458768 \\
            \hline
            \multicolumn{2}{l|}{hbmqtt\_\_two\_client} & $0$ & 243495 & 80928 & 9 & 32486 & 3193 & 9 & 32486 & 3193 \\
            $|\mathcal{M}| = 9$ & $|I| = 9$ & $1$ & 2916000 & 729000 & 25 & 89799 & 9220 & 23 & 85792 & 8285 \\
             & & $2$ & 32805000 & 6561000 & 66 & 252707 & 22341 & 59 & 233751 & 21405 \\
             & & $3$ & \xmark & \xmark & 173 & 714156 & 57517 & 153 & 647159 & 55546 \\
            \hline
        \end{tabular}
        }
    \caption{Benchmarking results for MQTT models (part 1)}
    \label{tab:raw_results_MQTT1}
\end{table}

\newpage

\begin{table}[!htbp]
    \centering
    {\scriptsize
        \begin{tabular}{p{1.2cm}p{1.8cm}|c|x{1.15cm}x{1.15cm}|x{0.7cm}x{1cm}x{1cm}|x{0.7cm}x{1cm}x{1cm}}
            & & & & & \multicolumn{6}{c}{Using expansion $\mathcal{M}^d$} \\
            \multicolumn{2}{l|}{Experiment} & & \multicolumn{2}{c|}{Adapted $L^*$} & \multicolumn{3}{c|}{\sout{Unique root assumption}} & \multicolumn{3}{c}{Unique root assumption} \\
            $|\mathcal{M}|$ & $|I|$ & $d$ & Inputs & Resets & $|\mathcal{M}^d|$ & Inputs & Resets & $|\mathcal{M}^d|$ & Inputs & Resets \\
            \hline\hline
            \multicolumn{2}{l|}{VerneMQ\_\_non\_clean} & $0$ & 312882 & 72033 & 12 & 48526 & 615 & 12 & 48526 & 615 \\
            $|\mathcal{M}| = 12$ & $|I| = 6$ & $1$ & 2304000 & 432000 & 35 & 174864 & 2017 & 32 & 174864 & 615 \\
             & & $2$ & 16416000 & 2592000 & 108 & 609195 & 6059 & 93 & 591195 & 615 \\
             & & $3$ & \xmark & \xmark & 336 & 2041269 & 18123 & 275 & 1927249 & 615 \\
            \hline
            \multicolumn{2}{l|}{hbmqtt\_\_two\_client\_} & $0$ & 544440 & 153343 & 17 & 94781 & 6849 & 17 & 94781 & 6849 \\
            \multicolumn{2}{l|}{will\_retain} & $1$ & 6237000 & 1377000 & 62 & 348748 & 25885 & 59 & 341298 & 24569 \\
            $|\mathcal{M}| = 17$ & $|I| = 9$ & $2$ & 68526000 & 12393000 & 218 & 1340374 & 89626 & 203 & 1275611 & 86860 \\
             & & $3$ & \xmark & \xmark & 772 & 5153270 & 328984 & 721 & 4967676 & 321186 \\
            \hline
            \multicolumn{2}{l|}{emqtt\_\_single\_client} & $0$ & 353302 & 109980 & 10 & 111130 & 24698 & 10 & 111130 & 24698 \\
            $|\mathcal{M}| = 10$ & $|I| = 11$ & $1$ & 5082000 & 1210000 & 27 & 405442 & 75698 & 27 & 405442 & 75698 \\
             & & $2$ & 69212000 & 13310000 & 78 & 1440586 & 228698 & 78 & 1440586 & 228698 \\
             & & $3$ & \xmark & \xmark & 231 & 5004594 & 687698 & 231 & 5004594 & 687698 \\
            \hline
            \multicolumn{2}{l|}{ActiveMQ\_\_single\_client} & $0$ & 253726 & 87923 & 8 & 75118 & 18503 & 8 & 75118 & 18503 \\
            $|\mathcal{M}| = 8$ & $|I| = 11$ & $1$ & 3751000 & 968000 & 19 & 250328 & 51503 & 19 & 250328 & 51503 \\
             & & $2$ & 51909000 & 10648000 & 48 & 796372 & 138503 & 48 & 796372 & 138503 \\
             & & $3$ & \xmark & \xmark & 124 & 2454372 & 366503 & 124 & 2454372 & 366503 \\
            \hline
            \multicolumn{2}{l|}{VerneMQ\_\_two\_client\_} & $0$ & 198945 & 76994 & 7 & 35583 & 919 & 7 & 35583 & 919 \\
            \multicolumn{2}{l|}{same\_id} & $1$ & 3025000 & 847000 & 25 & 141053 & 2082 & 19 & 131909 & 919 \\
            $|\mathcal{M}| = 7$ & $|I| = 11$ & $2$ & 42592000 & 9317000 & 86 & 505802 & 9207 & 53 & 409998 & 919 \\
             & & $3$ & \xmark & \xmark & 298 & 1838198 & 28524 & 145 & 1285009 & 919 \\
            \hline
            \multicolumn{2}{l|}{mosquitto\_\_two\_client} & $0$ & 504801 & 143865 & 16 & 85296 & 4257 & 16 & 85296 & 4257 \\
            $|\mathcal{M}| = 16$ & $|I| = 9$ & $1$ & 5832000 & 1296000 & 61 & 336310 & 18292 & 58 & 332278 & 17344 \\
             & & $2$ & 64152000 & 11664000 & 227 & 1352235 & 70385 & 215 & 1315305 & 67497 \\
             & & $3$ & \xmark & \xmark & 850 & 5404624 & 264603 & 806 & 5217721 & 258693 \\
            \hline
            \multicolumn{2}{l|}{mosquitto\_\_non\_clean} & $0$ & 312882 & 72033 & 12 & 48526 & 615 & 12 & 48526 & 615 \\
            $|\mathcal{M}| = 12$ & $|I| = 6$ & $1$ & 2304000 & 432000 & 35 & 174864 & 2017 & 32 & 174864 & 615 \\
             & & $2$ & 16416000 & 2592000 & 108 & 609195 & 6059 & 93 & 591195 & 615 \\
             & & $3$ & \xmark & \xmark & 336 & 2041269 & 18123 & 275 & 1927249 & 615 \\
            \hline
            \multicolumn{2}{l|}{emqtt\_\_simple} & $0$ & 42112 & 20986 & 3 & 17248 & 148 & 3 & 17248 & 148 \\
            $|\mathcal{M}| = 3$ & $|I| = 7$ & $1$ & 441000 & 147000 & 10 & 29352 & 1037 & 4 & 29262 & 148 \\
             & & $2$ & 4116000 & 1029000 & 20 & 98375 & 4055 & 5 & 44264 & 148 \\
             & & $3$ & 36015000 & 7203000 & 51 & 194336 & 7101 & 6 & 62264 & 148 \\
            \hline
            \multicolumn{2}{l|}{ActiveMQ\_\_non\_clean} & $0$ & 312882 & 72033 & 12 & 48526 & 615 & 12 & 48526 & 615 \\
            $|\mathcal{M}| = 12$ & $|I| = 6$ & $1$ & 2304000 & 432000 & 35 & 174864 & 2017 & 32 & 174864 & 615 \\
             & & $2$ & 16416000 & 2592000 & 108 & 609195 & 6059 & 93 & 591195 & 615 \\
             & & $3$ & \xmark & \xmark & 336 & 2041269 & 18123 & 275 & 1927249 & 615 \\
            \hline
            \multicolumn{2}{l|}{mosquitto\_\_mosquitto} & $0$ & 42112 & 20986 & 3 & 17248 & 148 & 3 & 17248 & 148 \\
            $|\mathcal{M}| = 3$ & $|I| = 7$ & $1$ & 441000 & 147000 & 10 & 29352 & 1037 & 4 & 29262 & 148 \\
             & & $2$ & 4116000 & 1029000 & 20 & 98375 & 4055 & 5 & 44264 & 148 \\
             & & $3$ & 36015000 & 7203000 & 51 & 194336 & 7101 & 6 & 62264 & 148 \\
            \hline
            \multicolumn{2}{l|}{mosquitto\_\_two\_client\_} & $0$ & 198945 & 76994 & 7 & 36561 & 919 & 7 & 36561 & 919 \\
            \multicolumn{2}{l|}{same\_id} & $1$ & 3025000 & 847000 & 25 & 146690 & 2082 & 19 & 137546 & 919 \\
            $|\mathcal{M}| = 7$ & $|I| = 11$ & $2$ & 42592000 & 9317000 & 86 & 518824 & 9207 & 53 & 418998 & 919 \\
             & & $3$ & \xmark & \xmark & 298 & 1930186 & 28524 & 145 & 1317003 & 919 \\
            \hline
            \multicolumn{2}{l|}{VerneMQ\_\_single\_client} & $0$ & 352869 & 109901 & 10 & 111768 & 24631 & 10 & 111768 & 24631 \\
            $|\mathcal{M}| = 10$ & $|I| = 11$ & $1$ & 5082000 & 1210000 & 27 & 408091 & 75631 & 27 & 408091 & 75631 \\
             & & $2$ & 69212000 & 13310000 & 78 & 1449166 & 228631 & 78 & 1449166 & 228631 \\
             & & $3$ & \xmark & \xmark & 231 & 5031166 & 687631 & 231 & 5031166 & 687631 \\
            \hline
            \multicolumn{2}{l|}{emqtt\_\_invalid} & $0$ & 66308 & 32978 & 3 & 32553 & 364 & 3 & 32553 & 364 \\
            $|\mathcal{M}| = 3$ & $|I| = 11$ & $1$ & 1089000 & 363000 & 17 & 86823 & 2082 & 5 & 68571 & 364 \\
             & & $2$ & 15972000 & 3993000 & 55 & 626885 & 14124 & 7 & 116573 & 364 \\
             & & $3$ & \xmark & \xmark & 261 & 1827065 & 38240 & 9 & 176573 & 364 \\
            \hline
            \multicolumn{2}{l|}{emqtt\_\_two\_client} & $0$ & 504801 & 143865 & 16 & 85296 & 4257 & 16 & 85296 & 4257 \\
            $|\mathcal{M}| = 16$ & $|I| = 9$ & $1$ & 5832000 & 1296000 & 61 & 336310 & 18292 & 58 & 332278 & 17344 \\
             & & $2$ & 64152000 & 11664000 & 227 & 1352235 & 70385 & 215 & 1315305 & 67497 \\
             & & $3$ & \xmark & \xmark & 850 & 5404624 & 264603 & 806 & 5217721 & 258693 \\
            \hline
            \multicolumn{2}{l|}{hbmqtt\_\_simple} & $0$ & 70182 & 34972 & 5 & 19425 & 7001 & 5 & 19425 & 7001 \\
            $|\mathcal{M}| = 5$ & $|I| = 7$ & $1$ & 735000 & 245000 & 16 & 36543 & 9129 & 16 & 36543 & 9129 \\
             & & $2$ & 6860000 & 1715000 & 31 & 77647 & 13204 & 31 & 77647 & 13204 \\
             & & $3$ & 60025000 & 12005000 & 60 & 198720 & 31241 & 60 & 198720 & 31241 \\
            \hline
        \end{tabular}
    }
    \caption{Benchmarking results for MQTT models (part 2)}
    \label{tab:raw_results_MQTT2}
\end{table}

\newpage

\begin{table}[!htbp]
    \centering
        {\scriptsize
        \begin{tabular}{p{1.2cm}p{1.5cm}|c|x{1.15cm}x{1.15cm}|x{0.7cm}x{1.15cm}x{1cm}|x{0.7cm}x{1.15cm}x{1cm}}
            & & & & & \multicolumn{6}{c}{Using expansion $\mathcal{M}^d$} \\
            \multicolumn{2}{l|}{Experiment} & & \multicolumn{2}{c|}{Adapted $L^*$} & \multicolumn{3}{c|}{\sout{Unique root assumption}} & \multicolumn{3}{c}{Unique root assumption} \\
            $|\mathcal{M}|$ & $|I|$ & $d$ & Inputs & Resets & $|\mathcal{M}^d|$ & Inputs & Resets & $|\mathcal{M}^d|$ & Inputs & Resets \\
            \hline\hline
            \multicolumn{2}{l|}{emqtt\_\_two\_client\_} & $0$ & 198945 & 76994 & 7 & 35583 & 919 & 7 & 35583 & 919 \\
            \multicolumn{2}{l|}{same\_id} & $1$ & 3025000 & 847000 & 25 & 141053 & 2082 & 19 & 131909 & 919 \\
            $|\mathcal{M}| = 7$ & $|I| = 11$ & $2$ & 42592000 & 9317000 & 86 & 505802 & 9207 & 53 & 409998 & 919 \\
             & & $3$ & \xmark & \xmark & 298 & 1838198 & 28524 & 145 & 1285009 & 919 \\
            \hline
            \multicolumn{2}{l|}{mosquitto\_\_two\_client\_} & $0$ & 652624 & 162340 & 18 & 163290 & 2176 & 18 & 163290 & 2176 \\
            \multicolumn{2}{l|}{will\_retain} & $1$ & 6804000 & 1458000 & 103 & 898463 & 3881 & 96 & 898463 & 2176 \\
            $|\mathcal{M}| = 18$ & $|I| = 9$ & $2$ & 74358000 & 13122000 & 513 & 4611829 & 14137 & 457 & 4519893 & 3162 \\
             & & $3$ & \xmark & \xmark & 2589 & 24227465 & 55352 & 2235 & 23406857 & 29150 \\
            \hline
        \end{tabular}}
    \caption{Benchmarking results for MQTT models (part 3)}
    \label{tab:raw_results_MQTT3}
\end{table}

\begin{table}[!htbp]
    \centering
        {\scriptsize
        \begin{tabular}{p{1.2cm}p{1.5cm}|c|x{1.15cm}x{1.15cm}|x{0.7cm}x{1.15cm}x{1cm}|x{0.7cm}x{1.15cm}x{1cm}}
            & & & & & \multicolumn{6}{c}{Using expansion $\mathcal{M}^d$} \\
            \multicolumn{2}{l|}{Experiment} & & \multicolumn{2}{c|}{Adapted $L^*$} & \multicolumn{3}{c|}{\sout{Unique root assumption}} & \multicolumn{3}{c}{Unique root assumption} \\
            $|\mathcal{M}|$ & $|I|$ & $d$ & Inputs & Resets & $|\mathcal{M}^d|$ & Inputs & Resets & $|\mathcal{M}^d|$ & Inputs & Resets \\
            \hline\hline
            \multicolumn{2}{l|}{GnuTLS\_3} & $0$ & 298250 & 88149 & 11 & 188005 & 52076 & 11 & 188005 & 52076 \\
            $|\mathcal{M}| = 11$ & $|I| = 8$ & $1$ & 3072000 & 704000 & 15 & 284005 & 73076 & 15 & 284005 & 73076 \\
             & & $2$ & 30208000 & 5632000 & 18 & 365005 & 89076 & 18 & 365005 & 89076 \\
             & & $3$ & \xmark & \xmark & 20 & 426005 & 100076 & 20 & 426005 & 100076 \\
            \hline
            \multicolumn{2}{l|}{OpenSSL\_1} & $0$ & 119182 & 41965 & 6 & 81361 & 27015 & 6 & 81361 & 27015 \\
            $|\mathcal{M}| = 6$ & $|I| = 7$ & $1$ & 1127000 & 294000 & 6 & 81361 & 27015 & 6 & 81361 & 27015 \\
             & & $2$ & 9947000 & 2058000 & 6 & 81361 & 27015 & 6 & 81361 & 27015 \\
             & & $3$ & 84035000 & 14406000 & 6 & 81361 & 27015 & 6 & 81361 & 27015 \\
            \hline
            \multicolumn{2}{l|}{GnuTLS\_3} & $0$ & 160296 & 55952 & 7 & 107552 & 35029 & 7 & 107552 & 35029 \\
            $|\mathcal{M}| = 7$ & $|I| = 8$ & $1$ & 1728000 & 448000 & 8 & 131555 & 41029 & 8 & 131555 & 41029 \\
             & & $2$ & 17408000 & 3584000 & 9 & 161556 & 47029 & 9 & 161556 & 47029 \\
             & & $3$ & \xmark & \xmark & 10 & 196556 & 53029 & 10 & 196556 & 53029 \\
            \hline
            \multicolumn{2}{l|}{OpenSSL\_1} & $0$ & 259591 & 76983 & 11 & 194765 & 54030 & 11 & 194765 & 54030 \\
            $|\mathcal{M}| = 11$ & $|I| = 7$ & $1$ & 2352000 & 539000 & 11 & 194765 & 54030 & 11 & 194765 & 54030 \\
             & & $2$ & 20237000 & 3773000 & 11 & 194765 & 54030 & 11 & 194765 & 54030 \\
             & & $3$ & \xmark & \xmark & 11 & 194765 & 54030 & 11 & 194765 & 54030 \\
            \hline
            \multicolumn{2}{l|}{GnuTLS\_3} & $0$ & 160296 & 55952 & 7 & 107552 & 35029 & 7 & 107552 & 35029 \\
            $|\mathcal{M}| = 7$ & $|I| = 8$ & $1$ & 1728000 & 448000 & 8 & 131555 & 41029 & 8 & 131555 & 41029 \\
             & & $2$ & 17408000 & 3584000 & 9 & 161556 & 47029 & 9 & 161556 & 47029 \\
             & & $3$ & \xmark & \xmark & 10 & 196556 & 53029 & 10 & 196556 & 53029 \\
            \hline
            \multicolumn{2}{l|}{JSSE\_1} & $0$ & 248336 & 71936 & 9 & 164780 & 45056 & 9 & 164780 & 45056 \\
            $|\mathcal{M}| = 9$ & $|I| = 8$ & $1$ & 2560000 & 576000 & 11 & 223796 & 55056 & 11 & 223796 & 55056 \\
             & & $2$ & 25088000 & 4608000 & 15 & 349804 & 74056 & 15 & 349804 & 74056 \\
             & & $3$ & \xmark & \xmark & 20 & 550804 & 102056 & 20 & 550804 & 102056 \\
            \hline
            \multicolumn{2}{l|}{JSSE\_1} & $0$ & 240344 & 71936 & 9 & 158776 & 44063 & 9 & 158776 & 44063 \\
            $|\mathcal{M}| = 9$ & $|I| = 8$ & $1$ & 2496000 & 576000 & 12 & 236801 & 57063 & 12 & 236801 & 57063 \\
             & & $2$ & 24576000 & 4608000 & 18 & 423816 & 85063 & 18 & 423816 & 85063 \\
             & & $3$ & \xmark & \xmark & 26 & 753816 & 129063 & 26 & 753816 & 129063 \\
            \hline
            \multicolumn{2}{l|}{OpenSSL\_1} & $0$ & 245252 & 69937 & 10 & 178588 & 47033 & 10 & 178588 & 47033 \\
            $|\mathcal{M}| = 10$ & $|I| = 7$ & $1$ & 2205000 & 490000 & 11 & 198588 & 52033 & 11 & 198588 & 52033 \\
             & & $2$ & 18865000 & 3430000 & 12 & 223588 & 57033 & 12 & 223588 & 57033 \\
             & & $3$ & \xmark & \xmark & 13 & 265588 & 64033 & 13 & 265588 & 64033 \\
            \hline
            \multicolumn{2}{l|}{OpenSSL\_1} & $0$ & 260877 & 89952 & 9 & 199182 & 65030 & 9 & 199182 & 65030 \\
            $|\mathcal{M}| = 9$ & $|I| = 10$ & $1$ & 3500000 & 900000 & 10 & 239185 & 73030 & 10 & 239185 & 73030 \\
             & & $2$ & 44000000 & 9000000 & 11 & 292186 & 82030 & 11 & 292186 & 82030 \\
             & & $3$ & \xmark & \xmark & 11 & 292186 & 82030 & 11 & 292186 & 82030 \\
            \hline
            \multicolumn{2}{l|}{NSS\_3} & $0$ & 200320 & 63944 & 8 & 147626 & 42007 & 8 & 147626 & 42007 \\
            $|\mathcal{M}| = 8$ & $|I| = 8$ & $1$ & 2112000 & 512000 & 8 & 147626 & 42007 & 8 & 147626 & 42007 \\
             & & $2$ & 20992000 & 4096000 & 8 & 147626 & 42007 & 8 & 147626 & 42007 \\
             & & $3$ & \xmark & \xmark & 8 & 147626 & 42007 & 8 & 147626 & 42007 \\
            \hline
            \multicolumn{2}{l|}{GnuTLS\_3} & $0$ & 313275 & 107944 & 9 & 246647 & 80033 & 9 & 246647 & 80033 \\
            $|\mathcal{M}| = 9$ & $|I| = 12$ & $1$ & 5040000 & 1296000 & 10 & 296650 & 90033 & 10 & 296650 & 90033 \\
             & & $2$ & 76032000 & 15552000 & 11 & 361651 & 101033 & 11 & 361651 & 101033 \\
             & & $3$ & \xmark & \xmark & 11 & 361651 & 101033 & 11 & 361651 & 101033 \\
            \hline
        \end{tabular}
    }
    \caption{Benchmarking results for TLS (part 1)}
    \label{tab:raw_results_TLS1}
\end{table}

\newpage

\begin{table}[!htbp]
    \centering
    {\scriptsize
        \begin{tabular}{p{1.2cm}p{1.8cm}|c|x{1.15cm}x{1.15cm}|x{0.7cm}x{1cm}x{1cm}|x{0.7cm}x{1cm}x{1cm}}
            & & & & & \multicolumn{6}{c}{Using expansion $\mathcal{M}^d$} \\
            \multicolumn{2}{l|}{Experiment} & & \multicolumn{2}{c|}{Adapted $L^*$} & \multicolumn{3}{c|}{\sout{Unique root assumption}} & \multicolumn{3}{c}{Unique root assumption} \\
            \hline\hline
            \multicolumn{2}{l|}{GnuTLS\_3} & $0$ & 321319 & 96042 & 12 & 204444 & 56099 & 12 & 204444 & 56099 \\
            $|\mathcal{M}| = 12$ & $|I| = 8$ & $1$ & 3328000 & 768000 & 18 & 348458 & 87099 & 18 & 348458 & 87099 \\
             & & $2$ & 32768000 & 6144000 & 26 & 606462 & 136099 & 26 & 606462 & 136099 \\
             & & $3$ & \xmark & \xmark & 32 & 842464 & 176099 & 32 & 842464 & 176099 \\
            \hline
            \multicolumn{2}{l|}{OpenSSL\_1} & $0$ & 413914 & 111985 & 16 & 309109 & 78055 & 16 & 309109 & 78055 \\
            $|\mathcal{M}| = 16$ & $|I| = 7$ & $1$ & 3675000 & 784000 & 18 & 359109 & 89055 & 18 & 359109 & 89055 \\
             & & $2$ & 31213000 & 5488000 & 19 & 389109 & 95055 & 19 & 389109 & 95055 \\
             & & $3$ & \xmark & \xmark & 19 & 389109 & 95055 & 19 & 389109 & 95055 \\
            \hline
            \multicolumn{2}{l|}{NSS\_3} & $0$ & 184272 & 55952 & 7 & 141522 & 41011 & 7 & 141522 & 41011 \\
            $|\mathcal{M}| = 7$ & $|I| = 8$ & $1$ & 1920000 & 448000 & 7 & 141522 & 41011 & 7 & 141522 & 41011 \\
             & & $2$ & 18944000 & 3584000 & 7 & 141522 & 41011 & 7 & 141522 & 41011 \\
             & & $3$ & \xmark & \xmark & 7 & 141522 & 41011 & 7 & 141522 & 41011 \\
            \hline
            \multicolumn{2}{l|}{miTLS\_0} & $0$ & 136256 & 47960 & 6 & 105435 & 35006 & 6 & 105435 & 35006 \\
            $|\mathcal{M}| = 6$ & $|I| = 8$ & $1$ & 1472000 & 384000 & 6 & 105435 & 35006 & 6 & 105435 & 35006 \\
             & & $2$ & 14848000 & 3072000 & 6 & 105435 & 35006 & 6 & 105435 & 35006 \\
             & & $3$ & \xmark & \xmark & 6 & 105435 & 35006 & 6 & 105435 & 35006 \\
            \hline
            \multicolumn{2}{l|}{RSA\_BSAFE\_Java\_6} & $0$ & 136256 & 47960 & 6 & 105435 & 35006 & 6 & 105435 & 35006 \\
            $|\mathcal{M}| = 6$ & $|I| = 8$ & $1$ & 1472000 & 384000 & 6 & 105435 & 35006 & 6 & 105435 & 35006 \\
             & & $2$ & 14848000 & 3072000 & 6 & 105435 & 35006 & 6 & 105435 & 35006 \\
             & & $3$ & \xmark & \xmark & 6 & 105435 & 35006 & 6 & 105435 & 35006 \\
            \hline
            \multicolumn{2}{l|}{OpenSSL\_1} & $0$ & 217535 & 69984 & 10 & 157702 & 48028 & 10 & 157702 & 48028 \\
            $|\mathcal{M}| = 10$ & $|I| = 7$ & $1$ & 2009000 & 490000 & 10 & 157702 & 48028 & 10 & 157702 & 48028 \\
             & & $2$ & 17493000 & 3430000 & 10 & 157702 & 48028 & 10 & 157702 & 48028 \\
             & & $3$ & \xmark & \xmark & 10 & 157702 & 48028 & 10 & 157702 & 48028 \\
            \hline
            \multicolumn{2}{l|}{RSA\_BSAFE\_C\_4} & $0$ & 208376 & 71936 & 9 & 156630 & 51019 & 9 & 156630 & 51019 \\
            $|\mathcal{M}| = 9$ & $|I| = 8$ & $1$ & 2240000 & 576000 & 11 & 220634 & 66019 & 11 & 220634 & 66019 \\
             & & $2$ & 22528000 & 4608000 & 11 & 220634 & 66019 & 11 & 220634 & 66019 \\
             & & $3$ & \xmark & \xmark & 11 & 220634 & 66019 & 11 & 220634 & 66019 \\
            \hline
            \multicolumn{2}{l|}{GnuTLS\_3} & $0$ & 617894 & 180370 & 15 & 453122 & 125131 & 15 & 453122 & 125131 \\
            $|\mathcal{M}| = 15$ & $|I| = 12$ & $1$ & 9504000 & 2160000 & 23 & 789136 & 197131 & 23 & 789136 & 197131 \\
             & & $2$ & \xmark & \xmark & 32 & 1247139 & 281131 & 32 & 1247139 & 281131 \\
             & & $3$ & \xmark & \xmark & 39 & 1659139 & 348131 & 39 & 1659139 & 348131 \\
            \hline
            \multicolumn{2}{l|}{OpenSSL\_1} & $0$ & 133217 & 48958 & 7 & 94403 & 33015 & 7 & 94403 & 33015 \\
            $|\mathcal{M}| = 7$ & $|I| = 7$ & $1$ & 1274000 & 343000 & 7 & 94403 & 33015 & 7 & 94403 & 33015 \\
             & & $2$ & 11319000 & 2401000 & 7 & 94403 & 33015 & 7 & 94403 & 33015 \\
             & & $3$ & \xmark & \xmark & 7 & 94403 & 33015 & 7 & 94403 & 33015 \\
            \hline
            \multicolumn{2}{l|}{GnuTLS\_3} & $0$ & 325454 & 107992 & 9 & 249665 & 77049 & 9 & 249665 & 77049 \\
            $|\mathcal{M}| = 9$ & $|I| = 12$ & $1$ & 5184000 & 1296000 & 12 & 355674 & 105049 & 12 & 355674 & 105049 \\
             & & $2$ & 77760000 & 15552000 & 17 & 555678 & 155049 & 17 & 555678 & 155049 \\
             & & $3$ & \xmark & \xmark & 22 & 802678 & 205049 & 22 & 802678 & 205049 \\
            \hline
            \multicolumn{2}{l|}{GnuTLS\_3} & $0$ & 621536 & 176475 & 16 & 444206 & 117170 & 16 & 444206 & 117170 \\
            $|\mathcal{M}| = 16$ & $|I| = 11$ & $1$ & 8712000 & 1936000 & 28 & 881237 & 214170 & 28 & 881237 & 214170 \\
             & & $2$ & \xmark & \xmark & 48 & 1777256 & 391170 & 48 & 1777256 & 391170 \\
             & & $3$ & \xmark & \xmark & 72 & 2937260 & 608170 & 72 & 2937260 & 608170 \\
            \hline
            \multicolumn{2}{l|}{OpenSSL\_1} & $0$ & 119182 & 41965 & 6 & 81361 & 27015 & 6 & 81361 & 27015 \\
            $|\mathcal{M}| = 6$ & $|I| = 7$ & $1$ & 1127000 & 294000 & 6 & 81361 & 27015 & 6 & 81361 & 27015 \\
             & & $2$ & 9947000 & 2058000 & 6 & 81361 & 27015 & 6 & 81361 & 27015 \\
             & & $3$ & 84035000 & 14406000 & 6 & 81361 & 27015 & 6 & 81361 & 27015 \\
            \hline
            \multicolumn{2}{l|}{NSS\_3} & $0$ & 433527 & 131930 & 11 & 370928 & 108016 & 11 & 370928 & 108016 \\
            $|\mathcal{M}| = 11$ & $|I| = 12$ & $1$ & 6768000 & 1584000 & 12 & 425929 & 119016 & 12 & 425929 & 119016 \\
             & & $2$ & \xmark & \xmark & 13 & 491930 & 130016 & 13 & 491930 & 130016 \\
             & & $3$ & \xmark & \xmark & 14 & 560930 & 140016 & 14 & 560930 & 140016 \\
            \hline
            \multicolumn{2}{l|}{OpenSSL\_1} & $0$ & 119182 & 41965 & 6 & 81361 & 27015 & 6 & 81361 & 27015 \\
            $|\mathcal{M}| = 6$ & $|I| = 7$ & $1$ & 1127000 & 294000 & 6 & 81361 & 27015 & 6 & 81361 & 27015 \\
             & & $2$ & 9947000 & 2058000 & 6 & 81361 & 27015 & 6 & 81361 & 27015 \\
             & & $3$ & 84035000 & 14406000 & 6 & 81361 & 27015 & 6 & 81361 & 27015 \\
            \hline
        \end{tabular}
    }
    \caption{Benchmarking results for TLS (part 2)}
    \label{tab:raw_results_TLS2}
\end{table}

\end{document}